\documentclass{aa} 

\usepackage{graphicx}
\usepackage{txfonts}
\usepackage{hyperref}
\usepackage{url}
\usepackage{xcolor}
\hypersetup{
colorlinks,
linkcolor={red!80!black},
citecolor={blue!80!black},
urlcolor={blue!80!black}
}

\def\cm3{cm$^{-3}$}
\def\kms{km~s$^{-1}$}

\def\lsun{L$_{\odot}$}
\def\rsun{R$_{\odot}$}

\def\msun{M$_{\odot}$}

\def\one{\ts {\,\sc i}}
\def\two{\ts {\,\sc ii}}
\def\three{\ts {\,\sc iii}}
\def\four{\ts {\,\sc iv}}

\def\beq{\begin{equation}}
\def\eeq{\end{equation}}

\def\lesssim{\mathrel{\hbox{\rlap{\hbox{\lower4pt\hbox{$\sim$}}}\hbox{$<$}}}}
\def\gtrsim{\mathrel{\hbox{\rlap{\hbox{\lower4pt\hbox{$\sim$}}}\hbox{$>$}}}}

\def\one{{\,\sc i}}
\def\two{{\,\sc ii}}
\def\three{{\,\sc iii}}
\def\four{{\,\sc iv}}

\def\v1d{{\sc v1d}}

\def\cmfgen{{\sc cmfgen}}
\def\heracles{{\sc heracles}}

\def\ergs{erg\,s$^{-1}$}

\newcommand{\iso}[2]{\ensuremath{^{#1}\rm{#2}}}

\def\aj{AJ}

\def\apj{ApJ}
\def\apjs{ApJS}
\def\apjl{ApJL}
\def\aap{A\&A}

\def\mnras{MNRAS}
\def\nat{Nature}

\def\jqsrt{JQSRT}

\def\nifs{\iso{56}Ni}

\begin{document}

   \title{
   Super-luminous Type II supernovae powered by magnetars
   }

   \titlerunning{
   Type II supernovae powered by magnetars
   }

\author{Luc Dessart\inst{\ref{inst1}}
  \and
  Edouard Audit\inst{\ref{inst2}}
  }

\institute{Unidad Mixta Internacional Franco-Chilena de Astronom\'ia (CNRS UMI 3386),
    Departamento de Astronom\'ia, Universidad de Chile,
    Camino El Observatorio 1515, Las Condes, Santiago, Chile\label{inst1}
    \and
    Maison de la Simulation, CEA, CNRS, Universit\'e Paris-Sud, UVSQ,
    Universit\'e Paris-Saclay, 91191, Gif-sur-Yvette, France.\label{inst2}
  }

   \date{Received; accepted}

  \abstract{
Magnetar power is  believed to be at the origin of numerous
super-luminous supernovae (SNe) of Type Ic, arising from compact,
hydrogen-deficient, Wolf-Rayet type stars.
Here, we investigate the properties that magnetar power would have on
standard-energy SNe associated with 15--20\,\msun\ supergiant stars,
either red (RSG; extended) or blue (BSG; more compact).
We have used a combination of Eulerian gray radiation-hydrodynamics
and non-LTE steady-state radiative transfer to study their dynamical,
photometric, and spectroscopic properties.
Adopting magnetar fields of 1, 3.5, $7 \times 10^{14}$\,G and rotational
energies of 0.4, 1, and $3 \times 10^{51}$\,erg, we produce bolometric
light curves
with a broad maximum covering $50-150$\,d  and a magnitude of $10^{43}-10^{44}$\,\ergs.
The spectra at maximum light are analogous to those of standard SNe II-P but bluer.
Although the magnetar energy is channelled in equal proportion between SN kinetic energy
and SN luminosity, the latter may be boosted by a factor of $10-100$
compared to a standard SN II.
This influence breaks the observed relation between brightness and ejecta expansion rate
of standard Type II SNe.
Magnetar energy injection also delays recombination and may even cause re-ionization,
with a reversal in photospheric temperature and velocity.
Depositing the magnetar energy in a narrow mass shell at the ejecta base leads to the formation
of a dense shell at a few 1000\,\kms, which causes a light-curve bump
at the end of the photospheric phase.
Depositing this energy over a broad range of mass in the inner ejecta,
to mimic the effect of multi-dimensional fluid instabilities,
prevents the formation of a dense shell and produces an earlier-rising and smoother
light curve. The magnetar influence  on the SN radiation is generally not visible
prior to $20-30$\,d, during which one may discern a BSG from a RSG progenitor.
We propose a magnetar model for the super-luminous Type II SN OGLE-SN14-073.
  }


   \maketitle
%

\section{Introduction}

A large number of super-luminous supernovae (SLSNe)
of Type II show unambiguous
evidence for interaction with circum-stellar material (CSM;
\citealt{ofek_06gy_07,smith_06gy_07,stoll_10jl_11}).
Because of the large kinetic energy in standard SN ejecta,
the observed time-integrated bolometric
luminosity can be explained by invoking ejecta deceleration
by a massive and dense CSM. The absence of broad, SN-like,
lines at early times and the presence of narrow lines
over an extended period of time is unambiguous evidence that
such an interaction takes place.
Some Type II SNe, however, show no obvious
signature of interaction despite having a luminosity a factor of
$10-100$ larger than standard SNe II at maximum, for example,
 SN\,2008es or OGLE-SN14-073 \citep{gezari_08es_09,miller_08es_09,terreran_slsn2_17}.
What is  striking in these events is the presence of H\one\ lines
well beyond the time of maximum, with a spectral morphology that is
analogous to Type II-SNe during the photospheric phase \citep{gutierrez_pap1_17}.
 This property excludes a large amount of \iso{56}Ni as the power source of the
 light curve because in that case, the emitting layers at and beyond
 bolometric maximum are necessarily rich in intermediate mass elements (IMEs)
 and iron-group elements (IGEs), as obtained in Type II SN models
 produced by the pair-production instability in super-massive H-rich stars
 \citep{d13_pisn}.

  An alternative power source to the interaction between ejecta and CSM is the injection
  of energy from a magnetized fast-spinning compact remnant (i.e. a magnetar).
  Magnetars may be associated with a wide range of astrophysical events,
  including $\gamma$-ray bursts
  \citep{duncan_pm_92,usov_pm_92,thompson_pm_04,bucciantini+08,bucciantini_pm_09,metzger_pm_11,metzger_pm_15},
  super or even extremely luminous
  SNe \citep{KB10,woosley_pm_10,bersten_15lh_16,sukhbold_woosley_16},
  or SNe presenting anomalies in their light curves such as a double bump
  \citep{maeda_05bf_07,taddia_pm_17}.
  In the context of super-luminous SNe, a fundamental feature of magnetars
  is their ability to supply power with a large range
  of magnitudes and time scales without stringent requirements on the ejecta mass
  or composition. The envelope and core properties of the progenitor
  might take diverse combinations, in contrast to pair-instability SNe whose
  large production of \nifs\ can only occur in a progenitor of huge mass \citep{barkat_pisn_67}.

   Magnetar power has been invoked to explain the double-peak light curve of SN\,2005bf
(\citealt{maeda_05bf_07}; see also \citealt{taddia_pm_17}).
\citet{KB10}  demonstrated that some combination of
magnetar field strength and initial spin period (as well as ejecta mass and kinetic energy)
could explain the observations of  the Type II SN \,2008es  \citep{gezari_08es_09,miller_08es_09}
and the type Ic SN\,2007bi \citep{galyam_07bi_09}. More recently, a large number of
Type Ic SLSNe have been discovered and followed photometrically and spectroscopically
from early to late times. Their radiative properties favor a magnetar origin
in a massive Wolf-Rayet progenitor
\citep{inserra_slsn_13,nicholl_slsn_13,nicholl_slsn_14,jerkstrand_slsnic_17}.
In most of these SLSNe Ic, distinguishing between \nifs\ power and magnetar power
can in fact be done from a single spectrum taken around bolometric maximum \citep{d12_magnetar}.
While most  simulations have assumed spherical symmetry, a few radiation-hydrodynamics simulations
performed in two dimensions suggest the occurrence of strong Rayleigh-Taylor driven mixing
in the inner ejecta, affecting the thermal and density structures of a large fraction of the ejecta,
which may affect the emergent radiation \citep{chen_pm_2d_16,suzuki_pm_2d_17}.

    In this paper, we present results from numerical simulations of magnetar powered
SNe resulting from the explosion of a red-supergiant (RSG) or a blue-supergiant (BSG) star.
In the next section, we present the numerical setup for the radiation hydrodynamics
and for the radiative transfer calculations. We then present our results.
In Section~\ref{sect_edep}, we first discuss the influence of the adopted
energy-deposition profile. We then describe the impact of the chemical
stratification on the bolometric light curve (Section~\ref{sect_nfx}).
In Section~\ref{sect_prog}, we present the light curve differences obtained for the explosion
of a BSG and a RSG progenitor influenced by the same magnetar.
Using a given explosion model from a RSG progenitor, we explore the diversity of
light curves and ejecta properties resulting from a grid of magnetar fields and spin periods (Section~\ref{sect_grid}).
In Section~\ref{sect_obs}, we discuss our model results in the context of observations.
We present our conclusions in Section~\ref{sect_conc}.


\begin{table*}
\caption{Summary of model properties for the progenitor, the ejecta, the magnetar,
and some results at bolometric maximum obtained with the \heracles\ simulations.
\label{tab_sum}}
\begin{center}
\begin{tabular}{l@{\hspace{4mm}}c@{\hspace{4mm}}c@{\hspace{4mm}}
c@{\hspace{4mm}}c@{\hspace{4mm}}c@{\hspace{4mm}}c@{\hspace{4mm}}
c@{\hspace{4mm}}rc@{\hspace{4mm}}c@{\hspace{4mm}}
c@{\hspace{4mm}}c@{\hspace{4mm}}
}
\hline
Model                & $R_{\star}$    & $M_{\rm ej}$ &   $E_{\rm kin}$    &   $E_{\rm pm}$      & $B_{\rm pm}$  & $m_{\rm lim}$  & $dm_{\rm lim}$  & $t_{\rm pm}$
                          &    $t_{\rm peak}$ &  $L_{\rm peak}$  & $V_{\rm phot, peak}$ \\
                          & [\rsun]            &     [\msun]    &  [erg] &  [erg]  &    [G]  &    [\msun] & [\msun] & [d]    &       [d]               &     [erg\,s$^{-1}$] & [\kms] \\
\hline
RE0p4B3p5     &      501          &        11.9        &        0.92(51)      &    0.4(51)  & 3.5(14) &   1.0  & 4.0  & 19  &       99    & 1.10(43)   & 3540   \\
RE0p4B3p5d     &      501          &        11.9      &        0.92(51)     &     0.4(51)  & 3.5(14) &   0.1  & 0.0  &  19 &     160  & 1.14(43)    & 2560  \\
RE0p4B3p5x     &      501          &        11.9      &        0.92(51)     &     0.4(51) & 3.5(14) &   1.0  & 4.0  &  19 &     102  &  1.14(43)    & 3510 \\
\hline
RE0p4B1        &      501          &         11.9        &        0.92(51)      &    0.4(51)  & 1.0(14) &  1.0  & 4.0  & 234 &      126     & 9.46(42)  & 2960  \\
RE0p4B3p5     &      501          &        11.9        &        0.92(51)      &    0.4(51)  & 3.5(14) &   1.0  & 4.0  & 19 &        99    & 1.10(43)  & 3540    \\
RE0p4B7         &      501          &        11.9        &        0.92(51)      &    0.4(51)  & 7.0(14) &  1.0  & 4.0  &   5 &       86     & 7.06(42)   & 3400\\

RE1B1             &      501          &        11.9        &        0.92(51)      &    1.0(51)  & 1.0(14) & 1.0  & 4.0  &   94  &       125  &  2.84(43) &4030  \\
RE1B3p5         &      501          &        11.9        &        0.92(51)      &    1.0(51)  & 3.5(14) &  1.0  & 4.0  &   8 &        76   & 1.77(43)   & 4880 \\
RE1B7             &      501          &        11.9        &        0.92(51)      &    1.0(51)  & 7.0(14) &  1.0  & 4.0  &   2 &        89   & 9.30(42)     &3820  \\

RE3B1             &      501          &        11.9        &        0.92(51)      &    3.0(51)  & 1.0(14) &1.0  & 4.0  &    31 &         86   & 7.78(43)   & 5680    \\
RE3B3p5         &      501          &        11.9        &        0.92(51)      &    3.0(51)  & 3.5(14) &  1.0  & 4.0  &   2.6  &        51   & 2.86(43)    & 6200  \\
RE3B7             &      501          &        11.9        &        0.92(51)     &    3.0(51)  & 7.0(14) &   1.0  & 4.0  &  0.6 &      90    & 1.19(43)      & 4390   \\
\hline
RE0p4B3p5     &      501          &        11.9        &        0.92(51)      &    0.4(51)  & 3.5(14) &  1.0  & 4.0  &  19 &       99    & 1.10(43)  & 3540  \\
BE0p4B3p5        &      47          &        15.6       &         1.20(51)    &    0.4(51)  & 3.5(14) &   1.0  & 4.0  &  19 &      105   & 9.57(42)   & 3230  \\
\hline
RE0p4B4p5o &      501          &        11.9        &        0.92(51)      &    0.4(51)  & 4.5(14) &  2.5  & 5.0  & 12 &   85 &    9.62(42)   &  4060 \\
RE0p4B4p5os &      501          &        17.8        &        2.67(51)      &    0.4(51)  & 4.5(14) &  2.5  & 5.0  & 12 &   83 &    1.09(43)   &  4280 \\
\hline
\end{tabular}
\end{center}
\end{table*}

\section{Numerical setup}
\label{sect_setup}

   We used the Eulerian radiation hydrodynamics code \heracles\
\citep{gonzalez_heracles_07,vaytet_mg_11} to simulate the influence
of magnetar power on a small set of SN ejecta models.
We first describe the code assumptions and set up,
then the treatment of magnetar power,
followed by the progenitor models employed,
and finally the post-processing for spectral calculations.
A summary of properties from
our simulations (initial conditions and results) is given in Table~\ref{tab_sum}.

\subsection{Numerical approach with \heracles}

    For the radiation-hydrodynamics simulations with \heracles, we assume
    spherical symmetry.
    This choice is most likely inadequate
    to describe the complex geometry of this system,
    both on large scales (see, e.g., \citealt{burrows_MHD_07},
    \citealt{bucciantini_pm_09}, \citealt{moesta_15})
    and small scales \citep{chen_pm_2d_16,suzuki_pm_2d_17}.
    Spherical symmetry has
    been assumed so far in all simulations of magnetar-powered SN light curves
    and we make the same simplification for convenience for the time being.
     We also employ a gray approximation for the radiative transfer
    (one energy/frequency group). In the 1-D simulations of interacting SNe
    presented in \citet{D15_2n,D16_2n}, we used a multi-group approach because
    the radiation and the gas could be strongly out of equilibrium, which does not
    apply as much here.
    We adopt a simple equation of state that treats the gas as ideal with $\gamma=$\,5/3
    and a mean atomic weight $\bar{A}$ of 1.35. The thermal energy of the gas is a tiny fraction
    of the radiative energy so the neglect, for example, of changes in the level of
    excitation and ionization of the gas has little impact
    on either the radiative or dynamical properties.
    The changes in ionization are accounted for in the computation
    of the Rossland mean opacity so the electron scattering opacity, which is the main
    contribution to the total opacity, is accurately accounted for. The code distinguishes
    absorptive from scattering opacity. At each grid point, the absorptive opacity is
    obtained by subtracting the electron scattering contribution to the Rosseland mean opacity
    (see \citealt{D15_2n} for details on how we compute these opacities).
    We have opacity tables for up to five different compositions, characteristic of a massive star
    envelope at core collapse. Namely, we adopt the composition of the H-rich envelope,
    the He-rich shell, the O-rich shell, the Si-rich shell, and a shell dominated by iron group elements.
    For simplicity, most of the simulations presented here adopt a composition with
     $X_{\rm H}=$\,0.65, $X_{\rm He}=$\,0.33, and a solar composition for heavy elements.
     (again, this composition is only relevant for the computation of the opacities)

    To better resolve the ejecta at smaller radii, we use a grid with a constant
    spacing from the minimum radius at $\sim$\,10$^{13}$\,cm up to
    $R_{\rm t} = 5\times$10$^{14}$\,cm,
    and then switch to a grid with a constant spacing in the log up to 10$^{16}$\,cm.
    The grid is designed to have no sharp jump in spacing at $R_{\rm t}$.
    For the boundary conditions, we adopt for both the gas and the radiation
    a reflecting inner boundary and a free-flow outer boundary.
    The bolometric light curves that we extract from the \heracles\ simulations
    are computed using the total radiative flux at the outer boundary.
    The light travel time to the outer boundary at 10$^{16}$\,cm introduces
    a delay for the record of the emergent radiation,  by at most $3-4$\,d.

     Radioactive decay is ignored in the present work.

\subsection{Treatment of magnetar energy deposition}

   We use the formulation of \citet{KB10} for the magnetar power as a function of time
since magnetar birth:
\begin{equation}
\dot{e}_{\rm pm} = (E_{\rm pm} / t_{\rm pm}) \,  / \left( 1 + (t + \delta t_{\rm pm})/ t_{\rm pm} \right)^2    \,
\end{equation}
with
\begin{equation}
 t_{\rm pm} = \frac{6 I_{\rm pm} c^3}{B_{\rm pm}^2  R_{\rm pm}^6  \omega_{\rm pm}^2} \,\, ,
\end{equation}
and where
$E_{\rm pm}$, $B_{\rm pm}$, $R_{\rm pm}$, $I_{\rm pm}$ and $\omega_{\rm pm}$ are the initial
rotational energy, magnetic field, radius, moment of inertia, and angular velocity of the magnetar;
$\delta t_{\rm pm}$ is the elapsed time since magnetar birth at the start of the \heracles\ simulation;
and $c$ is the speed of light.
We adopt $I_{\rm pm}=10^{45}$\,g\,cm$^2$ and $R_{\rm pm}=10^6$\,cm.
We use here the special case of a magnetic dipole spin down.
At late times,  the magnetar power scales as $1 / (B^2 t^2)$.

The term $\delta t_{\rm pm}$ is about 10$^5$\,s in our \heracles\ simulations because
we do not start at the time of explosion. Instead, we start from an already existing ejecta
at about 1\,d after magnetar birth.
This corresponds to a time shortly before shock breakout in our RSG
progenitor, and well after shock breakout in out BSG model (see next section).

In our simulations, we ignore the magnetar energy deposited during the first 10$^5$\,s
that follow core bounce and the magnetar birth.
This energy corresponds to
\begin{equation}
E_{\rm neglected} = E_{\rm pm} \left( 1 - 1 / (1 + \delta t_{\rm pm}/ t_{\rm pm}) \right) \, .
\end{equation}
$E_{\rm neglected}$ increases with $E_{\rm pm}$ and $B_{\rm pm}$.
For example, it can be as much as $1.8 \times 10^{51}$\,erg for $E_{\rm pm} = 3 \times 10^{51}$\,erg
and   $B_{\rm pm} = 7 \times 10^{14}$\,G  (so about 60\% of  $E_{\rm pm}$),
while it is merely $1.7 \times 10^{48}$\,erg  for $E_{\rm pm} = 0.4 \times 10^{51}$\,erg
and $B_{\rm pm} = 10^{14}$\,G (so less than a per cent of  $E_{\rm pm}$).
This neglect impacts moderately the emergent radiation because the
energy deposited prior to one day is mostly used to boost the ejecta kinetic energy
while it is strongly degraded by ejecta expansion.

The energy-deposition scheme is presented in detail in Section~\ref{sect_edep}.

\subsection{Progenitor models and initial conditions for the \heracles\ simulations}
\label{sect_init_prog}

We use two different models for the initial conditions in \heracles.
The first model (m15mlt3  of \citealt{d13_sn2p}) derives from
a red-supergiant (RSG) model of a 15\,\msun\ star on the main sequence.
At core collapse, it corresponds to a star with a total mass of 14.1\,\msun,
a surface radius of 500\,\rsun, a luminosity of 64\,200\,\lsun.
For the corresponding explosion, the total ejecta mass is 11.9\,\msun\
and the explosion energy is $0.92 \times 10^{51}$\,erg (slightly different
from the value in \citealt{d13_sn2p} because of remapping on a lower resolution
Eulerian grid initially and trimming of the inner regions at small radii).
The \nifs\ yield is irrelevant since radioactive decay is ignored in this study.
The second model (lm18a7Ad of \citealt{DH10}) derives from a blue-supergiant (BSG) model
of a 18\,\msun\ star on the main sequence (evolved at a metallicity of 0.008 rather than solar).
At the onset of collapse, it corresponds to a star with a total mass of 17\,\msun,
a surface radius of 47\,\rsun, and a luminosity of 210\,000\,\lsun.
For the corresponding explosion, the total ejecta mass is 15.6\,\msun\
and the explosion energy is $1.20 \times 10^{51}$\,erg.
Without magnetar power, these models reproduce closely the observed properties of
the standard Type II-P SN\,1999em and the Type II-peculiar SN\,1987A (see
\citealt{d13_sn2p} and \citealt{DH10} for discussion and results).

These two models are used to cover the range
of radii for Type II supernova progenitors, since the envelope extent
is the fundamental characteristic that distinguishes the progenitors
of Type II-Plateau and Type II-Peculiar SNe. Different progenitor
main sequence masses and/or adopted wind mass loss rates would also
impact the resulting observables of our magnetar-powered Type II SNe.
This second aspect is left to a future study.

Because the \heracles\ code is Eulerian, we start from an already existing ejecta rather
than one at core bounce.
For the RSG progenitor, we take the model at 10$^5$\,s after the explosion trigger,
which corresponds to about one hour before shock breakout.
For the BSG progenitor we take the same starting time, but because of the reduced progenitor
radius (i.e., 50\,\rsun), this time is well after shock breakout.
Since we focus on relatively massive ejecta for which the rise time to bolometric maximum
 (excluding the initial shock breakout burst) is weeks to months, this initial offset has little
 impact (see also discussion in the previous section).

We also need to specify the conditions between
the outer edge of the progenitor/ejecta at 10$^5$\,s and the outer grid radius at 10$^{16}$\,cm.
For simplicity, we fill this volume with a low-density low-temperature (set to 2000\,K)
material, reflecting a stellar wind mass loss rate of 10$^{-6}$\,\msun\,yr$^{-1}$.
A constant wind velocity of 50\,\kms\ is used for the RSG progenitor model,
and 500\,\kms\ for the BSG progenitor model.\footnote{
Our BSG progenitor/explosion model is suitable for SN\,1987A but the adopted wind
mass loss rate is too large and thus not strictly appropriate
(see e.g. \citealt{chevalier_dwarkadas_95}).}
These wind velocities are approximate but have little impact
on the results discussed here. Indeed, the ejecta/wind interaction contributes
a few percent of the total bolometric luminosity.
This contribution persists until the ejecta/wind interaction crosses the outer
boundary of the Eulerian grid, at 125\,d (50\,d) in the simulations based on the
RSG (BSG) progenitor.

Our model nomenclature is to use prefix 'R' ('B') to refer to simulations based on the RSG (BSG)
models. We then append this prefix by values for the magnetar initial rotational energy
and magnetic field.
For example, simulation RE1B3p5 employs the RSG progenitor model, a magnetar initial
rotational energy of 10$^{51}$\,erg and a magnetic field strength of $3.5 \times 10^{14}$\,G.
In Sections~\ref{sect_edep} and \ref{sect_nfx},
we explore special cases for which we stitch an additional suffix (for example `x' or `d').

\begin{figure*}
 \includegraphics[width=0.5\hsize]{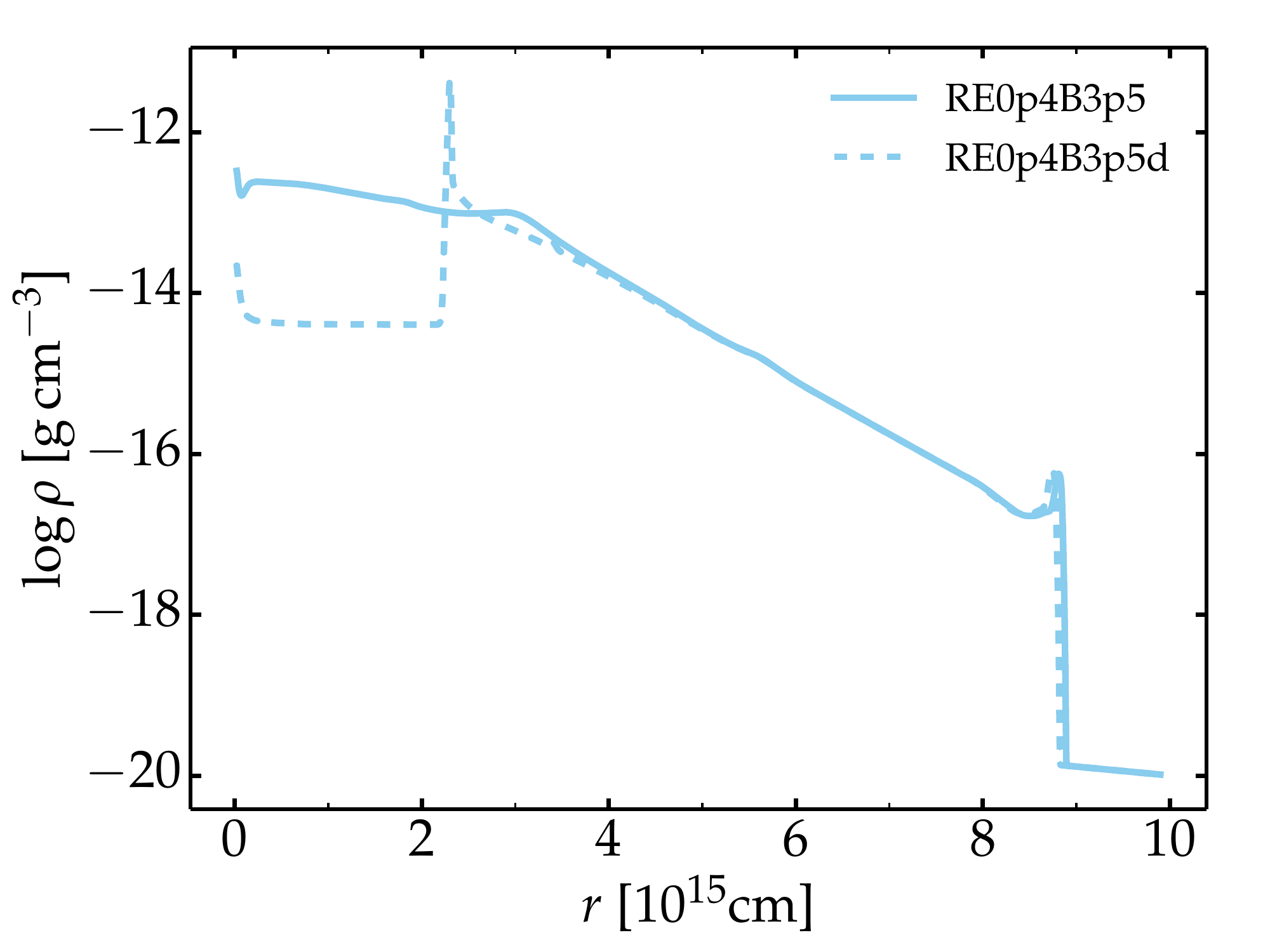}
 \includegraphics[width=0.5\hsize]{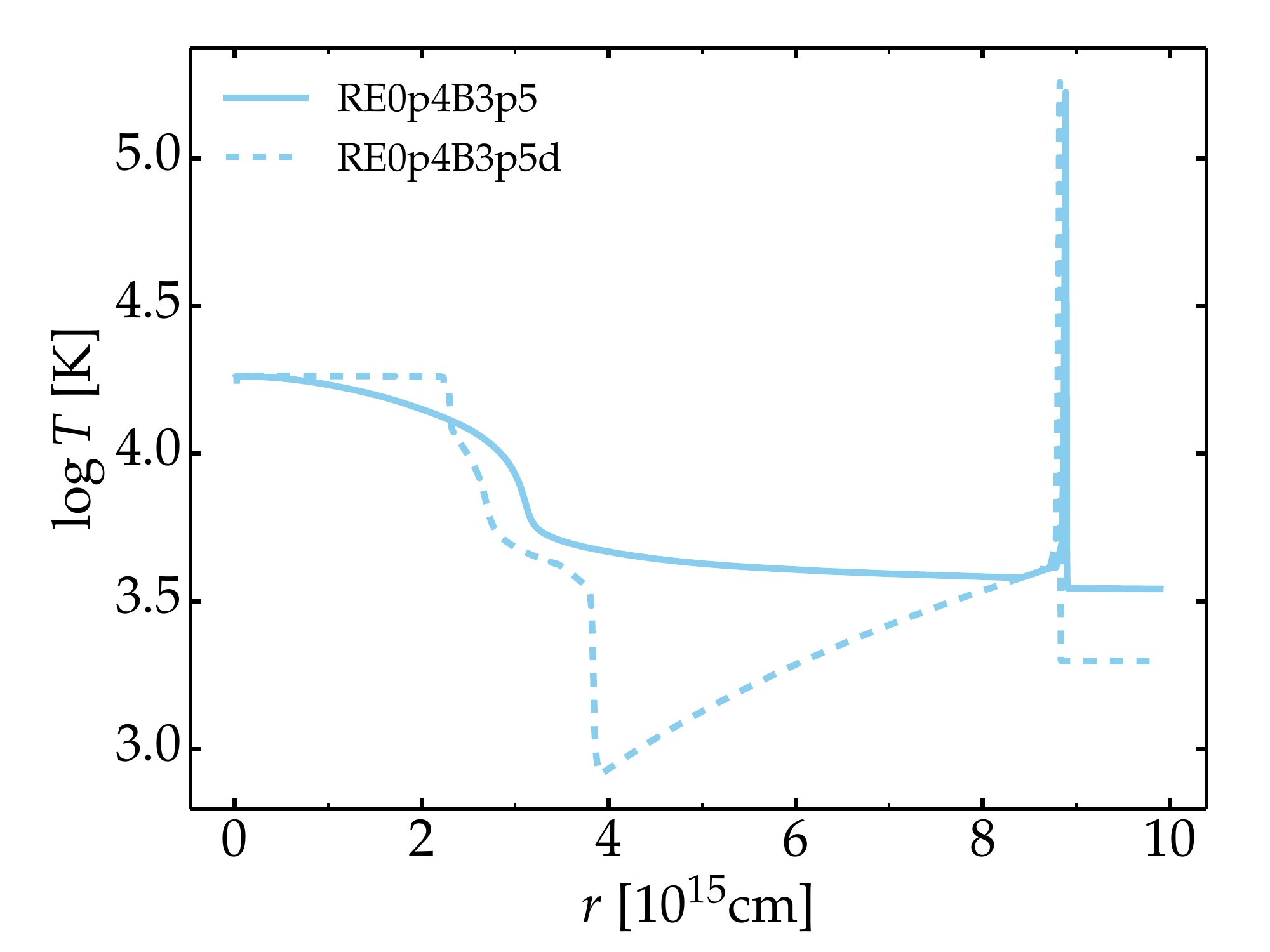}
 \includegraphics[width=0.5\hsize]{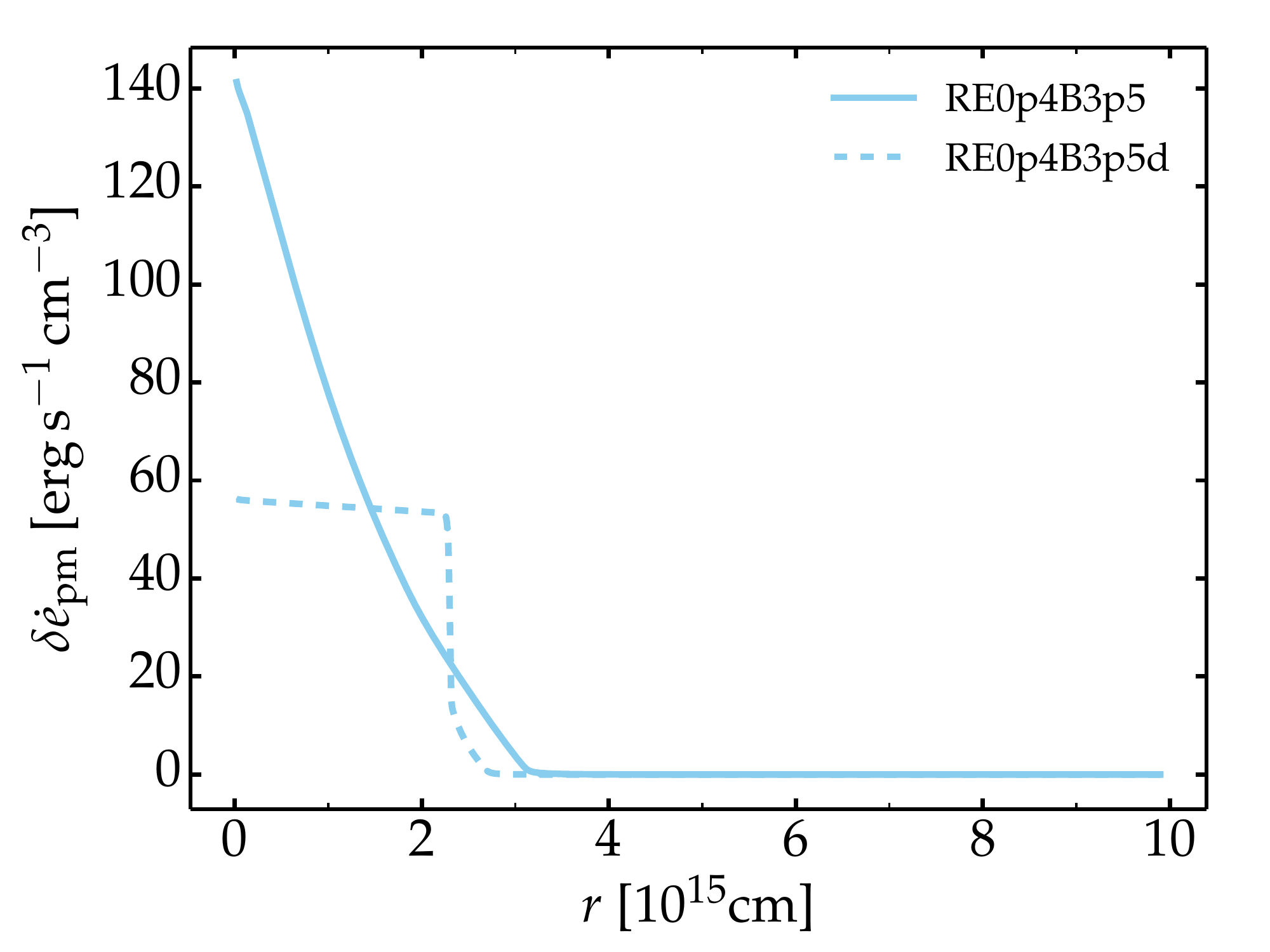}
 \includegraphics[width=0.5\hsize]{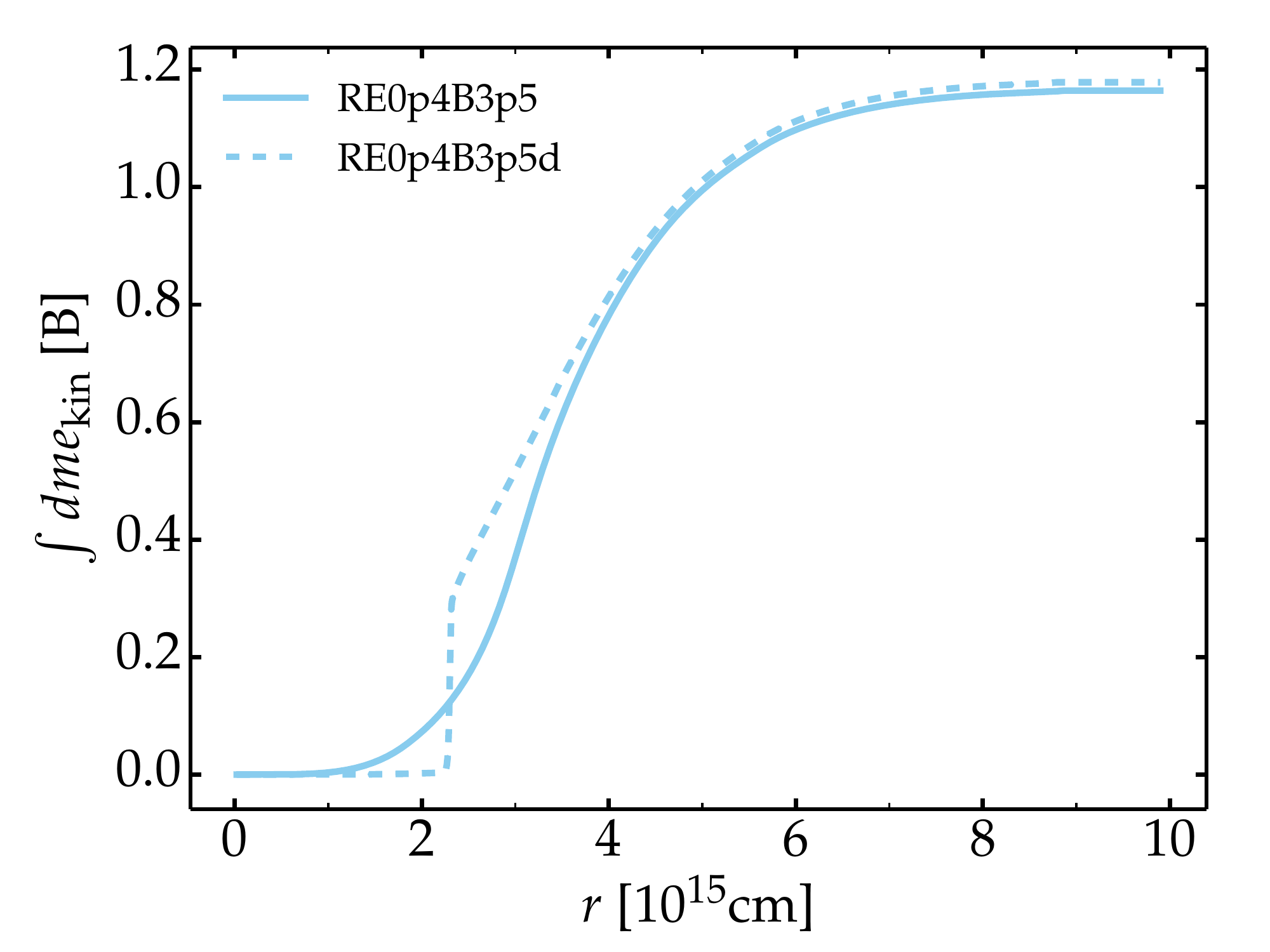}
   \caption{Comparison of models RE0p4B3p5 and RE0p4B3p5d
   to test the influence of the radial profile used for the deposition of magnetar power.
   Ordered clockwise from top left, we show the radial variation of the
   mass density, temperature, local emissivity from magnetar power,
   and the cumulative kinetic energy integrated outward
   from the inner boundary, all at 116\,d after the start of the \heracles\ simulation.
   Model RE0p4B3p5 spreads the magnetar energy over a large range of masses,
   while model RE0p4B3p5d deposits all the magnetar energy over the inner 0.1\,\msun.}
   \label{fig_pm_edep}
\end{figure*}

\begin{figure}
 \includegraphics[width=\hsize]{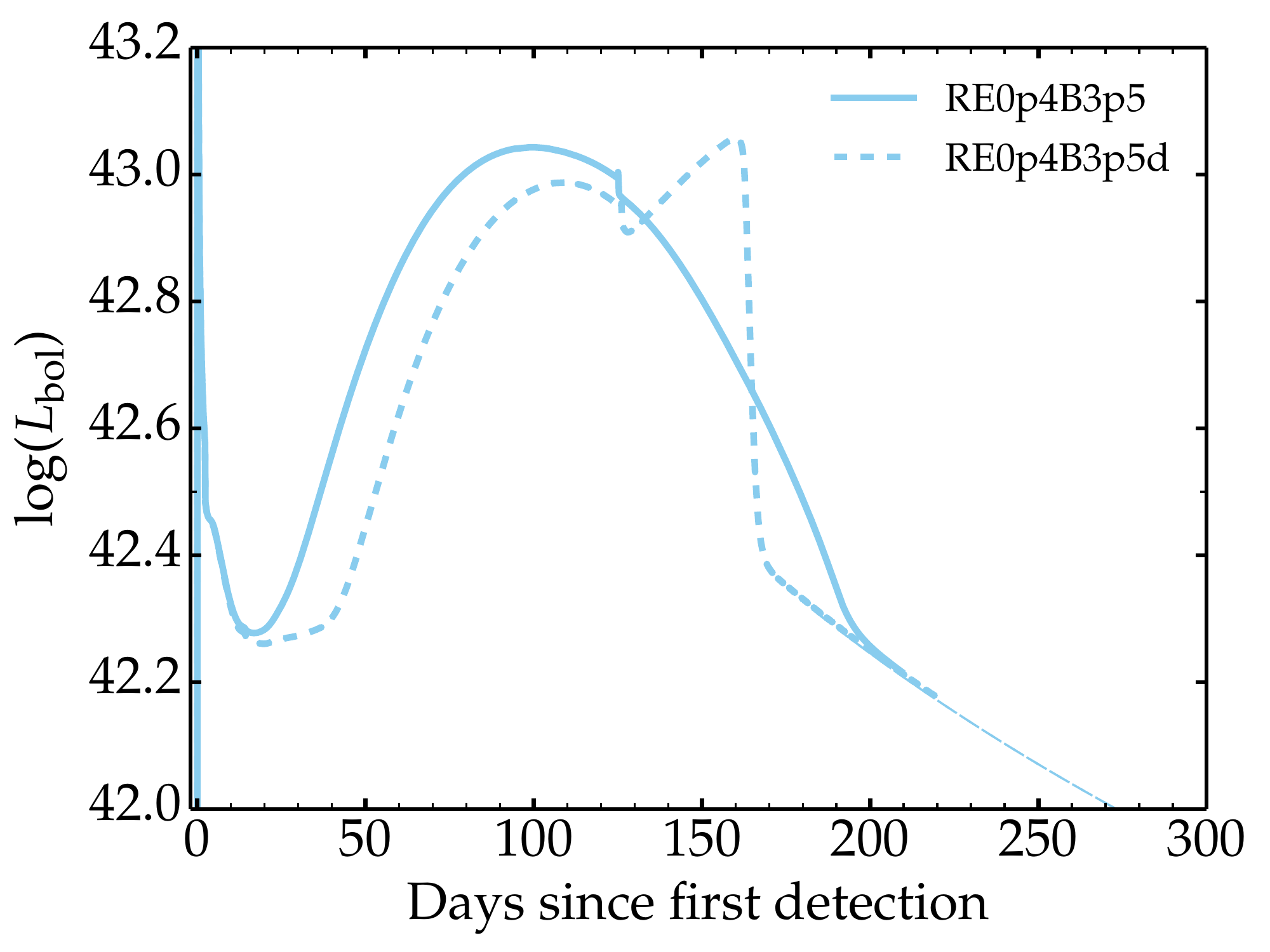}
   \caption{Comparison of the bolometric light curve for simulations
RE0p4B3p5 and RE0p4B3p5d, in which only the magnetar energy deposition
profile differs (see Fig.~\ref{fig_pm_edep}, and Section~\ref{sect_edep} for discussion).
The glitch at 125\,d occurs when the outer shock with the progenitor wind
leaves the Eulerian grid at 10$^{16}$\,cm.
The broad bump at $150-170$\,d for model RE0p4B3p5d occurs when the photosphere reaches
the dense shell in the inner ejecta; the sharp drop at 170\,d is when this
dense shell becomes optically thin.
   }
   \label{fig_lbol_edep}
\end{figure}

\subsection{Spectral simulations}

     We post-process the \heracles\ simulations at the time when the SN reaches bolometric
maximum, which is around 100\,d after the start of the simulation for our sample.
We use \cmfgen\ in a steady-state mode
and adopt the ejecta properties from \heracles\ (i.e., radius, velocity, density, temperature).
Homologous expansion is not assumed, which is why we read both the radius and the velocity.
In most simulations, we adopt a uniform composition of a RSG star at death.
We use mass fractions of 0.66 for H, 0.32 for He, 0.0019 for C, 0.004 for N, 0.008 for O,
and use the solar metallicity value for heavier elements.
The \cmfgen\ simulations treat H, He, C, N, O, Na, Si, Ca, Ti, Sc, and Fe.
The model atom includes H\one, He\one-\two, C\one--\three, N\one--\three, O\one--\three,
Na\one, Si\two--\three, Ca\two, Ti\two, Sc\two, and Fe\one--\four.
For the starting level populations in \cmfgen, we adopt local thermodynamic equilibrium
and then evolve the non-LTE solution with the temperature fixed to what it was in \heracles.

Evolving \cmfgen\ at a fixed temperature is not  consistent but it provides
a useful estimate of the emergent spectral energy distribution as well as
the ionization, color, and line profile width/strength.
During the photospheric phase, the ejecta is optically thick and lines do not dominate the luminosity.
At late times, when the ejecta is optically thin, lines become the main coolant so
that the LTE conditions assumed for the gas state in \heracles\ are no longer
suitable for a post-processing with \cmfgen. The \cmfgen\ simulations presented
here are therefore limited to the photospheric phase. We use the (unambiguous)
time of maximum for our comparisons.


\section{Influence of energy deposition profile}
\label{sect_edep}

   We first start with the problematic issue of the treatment of magnetar energy
deposition. The newly-born hot magnetar probably emits high energy photons
in the X-ray and $\gamma$-ray range as well as leptons (electrons and positrons).
The lepton energy should be efficiently thermalized but high energy photons
have a much larger mean free path and may eventually deposit their energy
far from their production site (they may even escape).

Secondly, as expected and demonstrated in recent 2-D simulations
\citep{chen_pm_2d_16,suzuki_pm_2d_17}, even if one assumes that
the energy is deposited and thermalized at the ejecta base, this magnetar energy injection
gives rise to strong Rayleigh-Taylor mixing in the inner ejecta.
Rather than slowly diffusing out through the ejecta as
it expands, the magnetar energy is more rapidly advected out by turbulent motions.
In multiple dimensions, this turbulence prevents the formation of a fast moving dense
shell in the inner ejecta, whose occurrence is an artifact of assuming spherical symmetry
\citep{KB10, woosley_pm_10}.

In this work, to mimic the effects of multi-dimensional instabilities seen in the simulations
of \citet{chen_pm_2d_16} and \citet{suzuki_pm_2d_17},
as well as the non-local nature of energy deposition of the magnetar,
we deposit the magnetar
energy over a range of mass shells rather than at the base of the ejecta.
We use an energy deposition that has the same profile as the density, with
an additional weight set to unity below a certain mass limit $m_{\rm lim}$.
In ejecta mass shells $m$ beyond $m_{\rm lim}$, the weight is either set to zero or
to $\exp(-x^2)$, where $x = (m-m_{\rm lim}) / dm_{\rm lim}$ ($m$ is taken as zero
at the ejecta base).  The adopted energy deposition profile is fixed in mass space
throughout the simulation. This is clearly very simplistic but it will allow us
to gauge the impact on observables.
In the future, it will be desirable to improve on this by performing multi-group
radiative transfer (to solve for the transport of both high energy and low-energy photons),
coupled with
multi-dimensional hydrodynamics to describe adequately the contributions of energy
transport by advection, diffusion, and non-local energy deposition (from high
energy particles and photons with a large mean free path).

   To test the influence of the magnetar energy deposition profile, we have run two
simulations based on the RSG progenitor model and influenced by a magnetar
with an initial rotational energy of $0.4 \times 10^{51}$\,erg and a magnetar field strength
of $3.5 \times 10^{14}$\,G.
In model RE0p4B3p5d, the weight is set to unity below $m_{\rm lim}=0.1$\,\msun,
and to zero above, that is the energy is deposited over a narrow mass range of 0.1\,\msun\
above the ejecta base.
In model RE0p4B3p5, the weight is set to unity below $m_{\rm lim}=1.0$\,\msun,
and to $\exp(-x^2)$ above, with $dm_{\rm lim}=4.0$\,\msun.
The corresponding volume integrated emissivity is then normalized to the
magnetar power at the time.

We show a set of results for this test comparison in Fig.~\ref{fig_pm_edep},
including the density, the temperature, the energy-deposition profile, and the
cumulative kinetic energy versus radius (the time is 116\,d, or 10$^7$\,s, for each simulation).
In the case where the magnetar energy is deposited in the inner 0.1\,\msun\ of the ejecta,
the density profile exhibits a dense shell at $\sim 2.2 \times 10^{15}$\,cm and 2200\,\kms\
(the velocity profile, not shown, is essentially homologous for each simulation), with little
mass below it. The temperature profile shows much more structure, with sharp variations
in the cool regions above the photosphere. The temperature spike at large radii in both
simulations corresponds to the shock with the surrounding low-density wind --- the
interaction power is small in comparison to the magnetar power.

The impact on the light curve is significant (Fig.~\ref{fig_lbol_edep}).
The time-integrated luminosity is greater by $1.5 \times 10^{49}$\,erg
in the model RE0p4B3p5 characterized by a smooth/extended magnetar energy deposition.
This excess radiative energy in model RE0p4B3p5 yields instead an
excess kinetic energy in the model RE0p4B3p5d. This is a one per cent difference
since the total ejecta kinetic energy is $\lesssim$\,1.2$\times$\,10$^{51}$\,erg
(bottom right panel of Fig.~\ref{fig_pm_edep}).
The morphology of the light curve is also affected by the adopted treatment
of magnetar energy deposition.
Model  RE0p4B3p5d, in which the deposition is confined to the innermost ejecta layers
and causes the formation of a dense shell, the light curve shows a pronounced bump
at 150\,d, which corresponds to the epoch when the photosphere recedes to those deep
ejecta layers. Model  RE0p4B3p5,  in which the deposition is spread in mass space shows
a smooth bolometric light curve. In this case, the onset of brightening also occurs $\sim$\,20\,d sooner because of the energy deposition further out in the ejecta, at smaller optical depths.
The simulations of \citet{KB10} do not show any jump in the light curve despite the
formation of a dense shell at the base of their ejecta. This is probably because they adopt
a fixed opacity, independent of ionization. Our models are H rich and the opacity varies
steeply when H recombines, as occurs at 150\,d in the dense shell formed
in model RE0p4B3p5d.

While the energy deposition implemented in both models is artificial, the smooth
and extended deposition profile adopted for model RE0p4B3p5 yields ejecta properties
in better agreement with the 2-D simulations of \citet{chen_pm_2d_16} and
\citet{suzuki_pm_2d_17}, in particular with the lack of a dense shell in the inner ejecta.
A similar effect on the density structure and on the resulting SN light curve is seen
in radiation-hydrodynamics simulations based on 3D simulations of core-collapse SNe
\citep{utrobin_2p_17}.

In the remaining of this work, we employ the same parametrized energy deposition
profile as for RE0p4B3p5. Our magnetar-powered simulations will therefore tend to produce material
distributed smoothly in velocity space down to very small values.

\begin{figure}
   \includegraphics[width=\hsize]{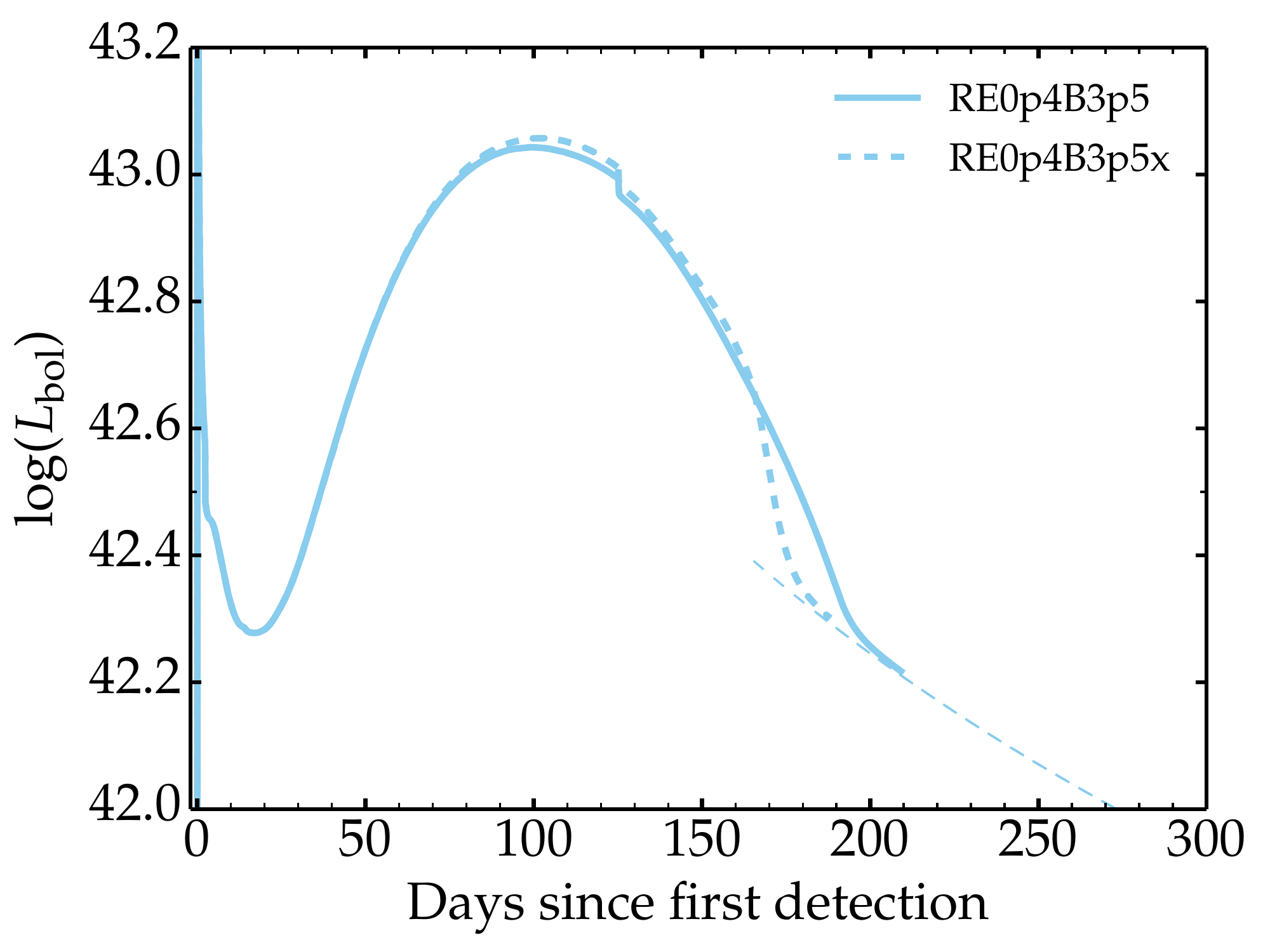}
   \caption{Bolometric light curves computed with \heracles\ for models
   RE0p4B3p5 (homogenous composition with $\bar{A} = 1.35$) and RE0p4B3p5x
   (chemically stratified and spanning $\bar{A} = 1.35$ up to 17).
   Allowing for a depth dependent abundance only affects the
   light curve after maximum, when the former (metal-rich) He-core material is progressively
   revealed. The impact on the light curve is however minor here.
   The glitch at 125\,d occurs when the outer shock with the progenitor wind
   leaves the Eulerian grid at 10$^{16}$\,cm.
   The spectra for these two models at maximum light are identical (hence not shown).
   This is because the spectrum then forms within the H-rich layers, and thus at the same composition
   in both models.
}
   \label{fig_pm_nfx}
\end{figure}


   \begin{figure*}
   \includegraphics[width=0.49\hsize]{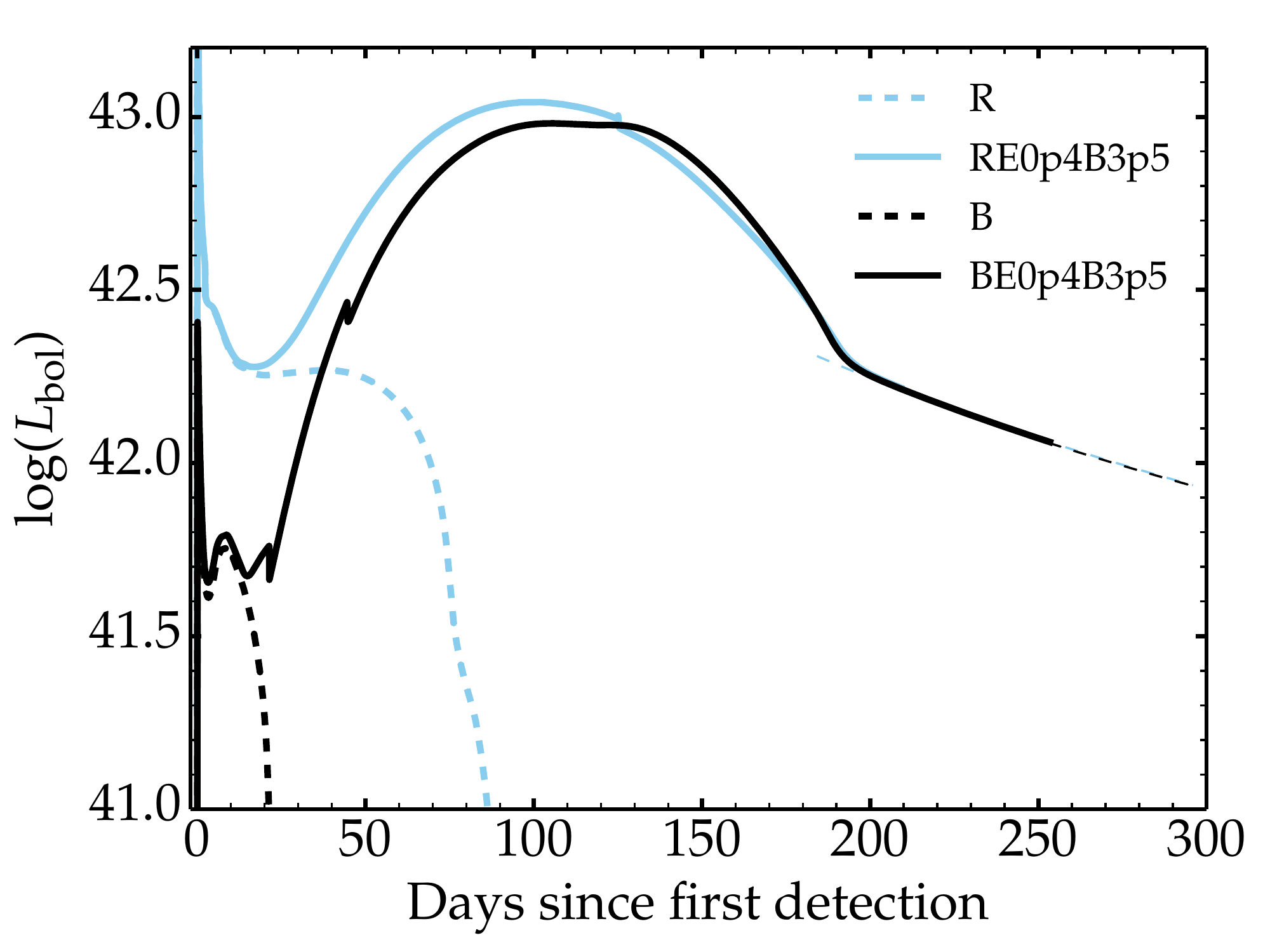}
   \includegraphics[width=0.49\hsize]{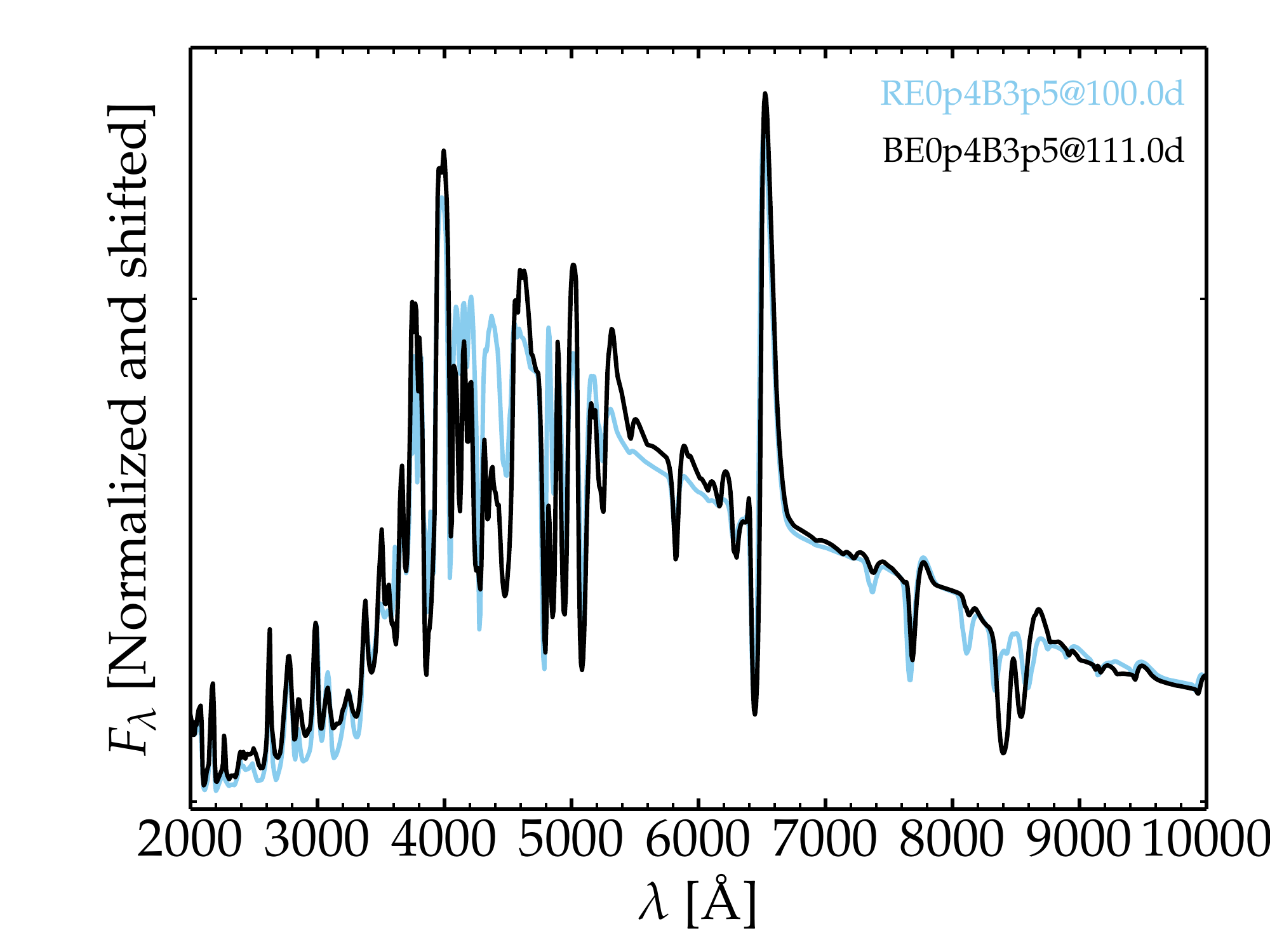}
   \caption{
   Left: Bolometric light curve computed with \heracles\ for
   models RE0p4B3p5 (RSG progenitor) and BE0p4B3p5 (BSG progenitor)
   under the influence of a magnetar (thick line) or not (thick dashed line).
   The thin dashed line corresponds to the magnetar power at $>$\,250\,d.
   Prior to the influence of magnetar power on the emergent light,
   the luminosity stems primarily from shock deposited energy.
   It is at such early times that one may distinguish a BSG from
   a RSG star as the progenitor of  a super-luminous SN.
   Right: Maximum light spectra (around 100\,d after explosion)
   for the two models  shown at left.
   }
   \label{fig_bsg_rsg}
\end{figure*}

\section{Influence of chemical stratification}
\label{sect_nfx}

   We also tested the influence of chemical stratification on our results with \heracles.
In \heracles, one can follow the evolution of ``scalars'' from their initial
distribution. In our Type II SN explosions, we approximate the ejecta composition
using only the five most important species. We focus on
the distribution of H, He, O, Si, and the initial \nifs\ (here, \nifs\ is just a tracer for IGEs since
we ignore radioactive decay). We renormalize the mass fraction
to unity to correct for the missing species. The only impact of this chemical stratification
on the numerical setup is through the opacity tables. Rather than using one table
for an H-rich composition, we now use five tables to track the evolution in mean
atomic weight $\bar{A}$ from the outer layers rich in H and He ($\bar{A} \approx 1.35$)
down to the O-Si rich layers ($\bar{A}  \approx 17.0$; this value depends on the progenitor/ejecta
composition and the adopted mixing).
We use the Rosseland mean opacity in our calculations so we account for the continuum opacity
from bound-free and free-free processes, and from electron scattering opacity, as well as line opacity.

We have tested the impact of chemical stratification using model RE0p4B3p5x --- model
RE0p4B3p5 has a uniform composition representative of the progenitor surface.
We find no difference in dynamical properties between
model RE0p4B3p5x and model RE0p4B3p5.
The gas density matters but not the precise distribution of this mass between different species/isotopes.
Since we use the same equation of state for both models, we neglect the change in energy release
from recombination and de-excitation of atoms/ions.
However, the thermal pressure is always a tiny fraction of the total pressure in SNe, and even more so
in magnetar-powered SNe.

Figure~\ref{fig_pm_nfx} shows the bolometric light curves for models RE0p4B3p5x and RE0p4B3p5.
The difference is negligible at early times, which is expected since more than half the ejecta is
made of the progenitor H-rich envelope.
At and after maximum, a small difference is seen, RE0p4B3p5x being first brighter and then fainter
than model RE0p4B3p5. We interpret this results as arising from the earlier release of trapped radiation
energy in model RE0p4B3p5x. The greater abundance of heavy elements in
model RE0p4B3p5x leads to a (modest) reduction in opacity, which makes
the trapping of radiation less efficient.
We use the Rosseland mean in our calculations so the reduction
in opacity is driven by the reduction in electron-scattering opacity,
and is not compensated by the greater metal line opacity at large $\bar{A}$.


\section{Influence of the initial structure and radius: BSG versus RSG progenitors}
\label{sect_prog}

   \citet{KB10} considered a variety of magnetar properties (magnetic
field and initial rotational energy), ejecta masses, and ejecta kinetic energies.
However, they neglect the initial internal energy left by the shock passage
(and that does not end up as kinetic energy for the ejecta), since this
energy is typically much less than the energy released by a magnetar
in a SLSN.
One consequence in their simulation is a low predicted luminosity at times shorter than
about the magnetar spin down time scale.
In reality, during this early phase, an important source of energy is the shock
deposited energy -- it is even the dominant source of energy in the explosion
of supergiant stars.
Indeed, in supergiant stars, this energy is large, not strongly degraded by expansion,
and allows even a standard SN II-P to radiate at 10$^{9}$\,\lsun\ for many days
after the shock emergence.
Accounting for the large size of supergiant progenitors is therefore necessary
to produce a consistent light curve prior to maximum (although the magnetar
energy deposition may extend far out in the ejecta and affect the SN brightness
very soon after explosion).

Here, we compare the predictions for the influence of a magnetar
($E_{\rm pm} = 0.4 \times 10^{51}$\,erg, $B_{\rm pm} = 3.5 \times 10^{14}$\,G)
acting on a SN ejecta that resulted from the explosion of a BSG and a RSG progenitor star.
As discussed in the Section~\ref{sect_init_prog}, these models yield properties similar
to SN\,1987A and SN\,1999em when evolved with no magnetar influence
\citep{DH10,d13_sn2p}.

   The left panel of Fig.~\ref{fig_bsg_rsg} presents the bolometric light curve
   of models BE0p4B3p5 and RE0p4B3p5. The thick dashed lines correspond to
   the resulting \heracles\ light curve with no magnetar and no \nifs\ (we ignore
   radioactive decay in this study).
   For up to $20-30$\,d, the model luminosity is not influenced by
   the magnetar in our setup, but beyond that, the light curves of models BE0p4B3p5 and RE0p4B3p5
   slowly converge and eventually overlap soon after maximum.
   The resulting light curves have a bell shape morphology.
   As expected, the original star size matters only
   at early times, while at late times, the power supply is so large that it overwhelms
   the slight differences that the two models may have had at explosion or at collapse.
   But the sizable differences at early times may allow to constrain the progenitor size,
   in the same fashion as for distinguishing SNe II-pec from SNe II-P (e.g., SN\,1987A
   from SN\,1999em, for which the differences at early times are well documented).

   The right panel of Fig.~\ref{fig_bsg_rsg} shows the optical spectra for each model
   around the time of bolometric maximum.
   The differences are marginal. It is important to notice
   that in such magnetar-powered SNe from H-rich stars, the spectra at maximum light
   show a typical Type II spectrum. In a model whose bolometric maximum is powered
   by the decay from a large mass of \nifs, the optical spectra at bolometric maximum tend
   to show lines from intermediate mass elements, although H\one\ lines may still be seen
   if the progenitor radius is huge \citep{d13_pisn}.


\section{Results for a grid of RSG-star explosions influenced by magnetar power.}
\label{sect_grid}

   In this section, we present results for a grid of simulations in which the magnetar
   properties are varied. Using the RSG progenitor model, we cover magnetar
   initial rotational energies of 0.4, 1, and $3.5 \times 10^{51}$\,erg (corresponding
   for our adopted neutron star to initial spin periods of 7.0, 4.4, and 2.6\,ms)
   and magnetic field strength of 1, 3.5, and $7 \times 10^{14}$\,G. The magnetar
   spin down timescale covers from a day to several months (Table~\ref{tab_sum};
   $t_{\rm pm}$ scales as the inverse of $B_{\rm pm}^2 E_{\rm pm}$).

\subsection{Bolometric light curves}

\begin{figure}
\includegraphics[width=\hsize]{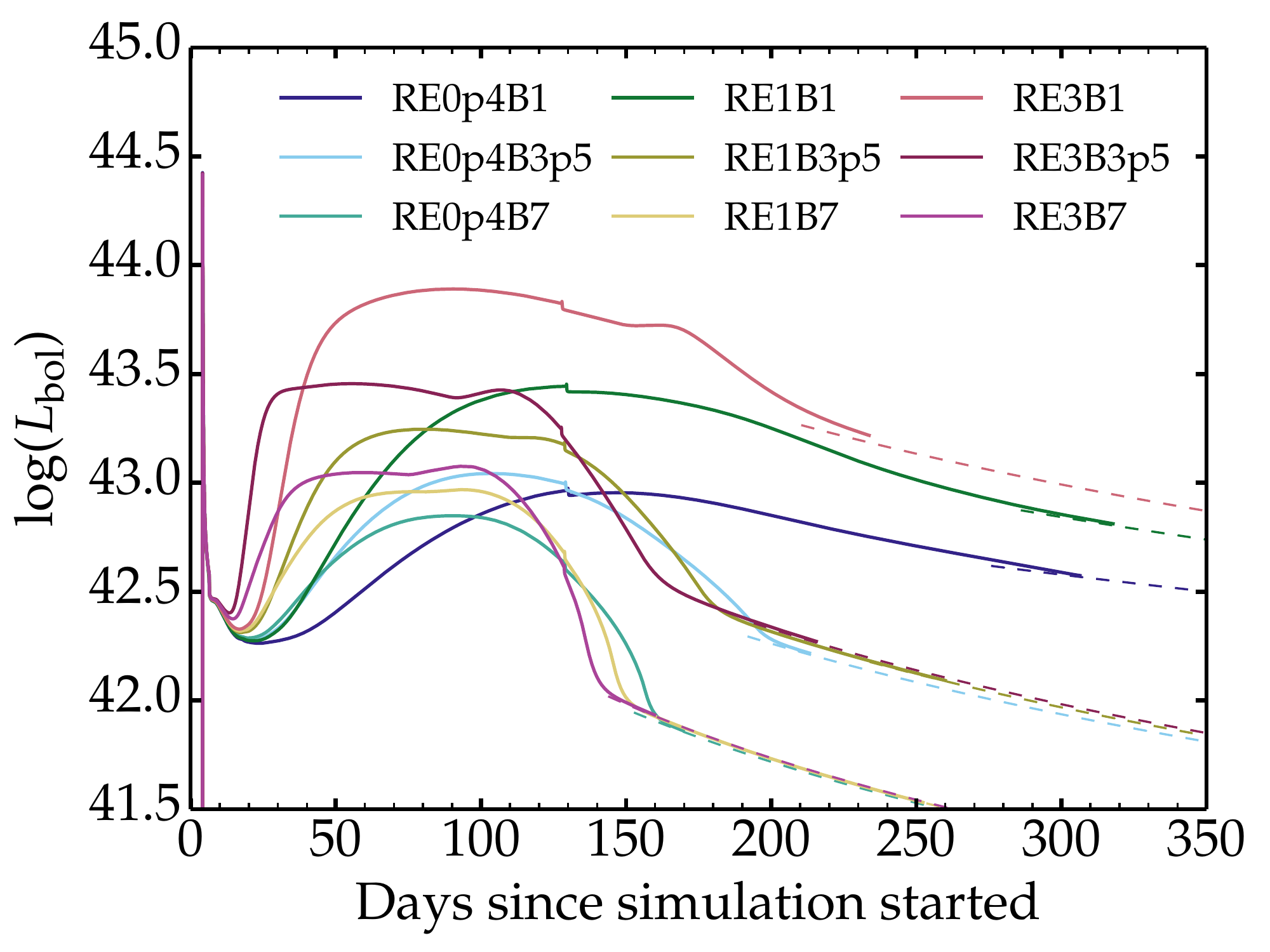}
\caption{
Bolometric light curves for the grid of magnetar-powered SN models
based on the RSG progenitor.
The glitch at 125\,d occurs when the outer shock with the progenitor wind
leaves the Eulerian grid at 10$^{16}$\,cm.
In the appendix, we present light curves for model sets sharing the same
$E_{\rm pm}$ or the same $B_{\rm pm}$.
(see Section~\ref{sect_grid} for discussion.)
}
\label{fig_pm_rsg}
\end{figure}

    We show the bolometric light curves for this set of models in Fig.~\ref{fig_pm_rsg}.
The light curves all start at the same level, when the magnetar influence  has not
yet been felt in the outer ejecta. This influence occurs sooner for a shorter spin down
timescale $t_{\rm pm}$. This depends also on the magnetar energy $E_{\rm pm}$
in our simulations but quantitatively, we need to be cautious since the energy released
prior to 1\,d is not accounted for in our simulations (this impacts our results for
small values of $t_{\rm pm}$).
The earlier the bolometric light curve rises again, the faster is the rise to the
bolometric maximum, which occurs in our set of simulations between 50 and 125\,d
after explosion.
The bolometric maximum spans the range $0.7-7.8 \times 10^{43}$\,erg\,s$^{-1}$.
These values for $t_{\rm peak}$ and $L_{\rm peak}$  are
broadly consistent\footnote{The broad and flat light curve maxima
can complicate an accurate estimate of the time of maximum.} with
the values obtained from the analytic expressions (15) and (16) in \citet{KB10}.
Beyond maximum, the bolometric luminosity falls onto the instantaneous magnetar
luminosity (which then scales as the inverse of $B_{\rm pm}^2$) after a time
similar to $t_{\rm peak}$.

\begin{figure}
\includegraphics[width=\hsize]{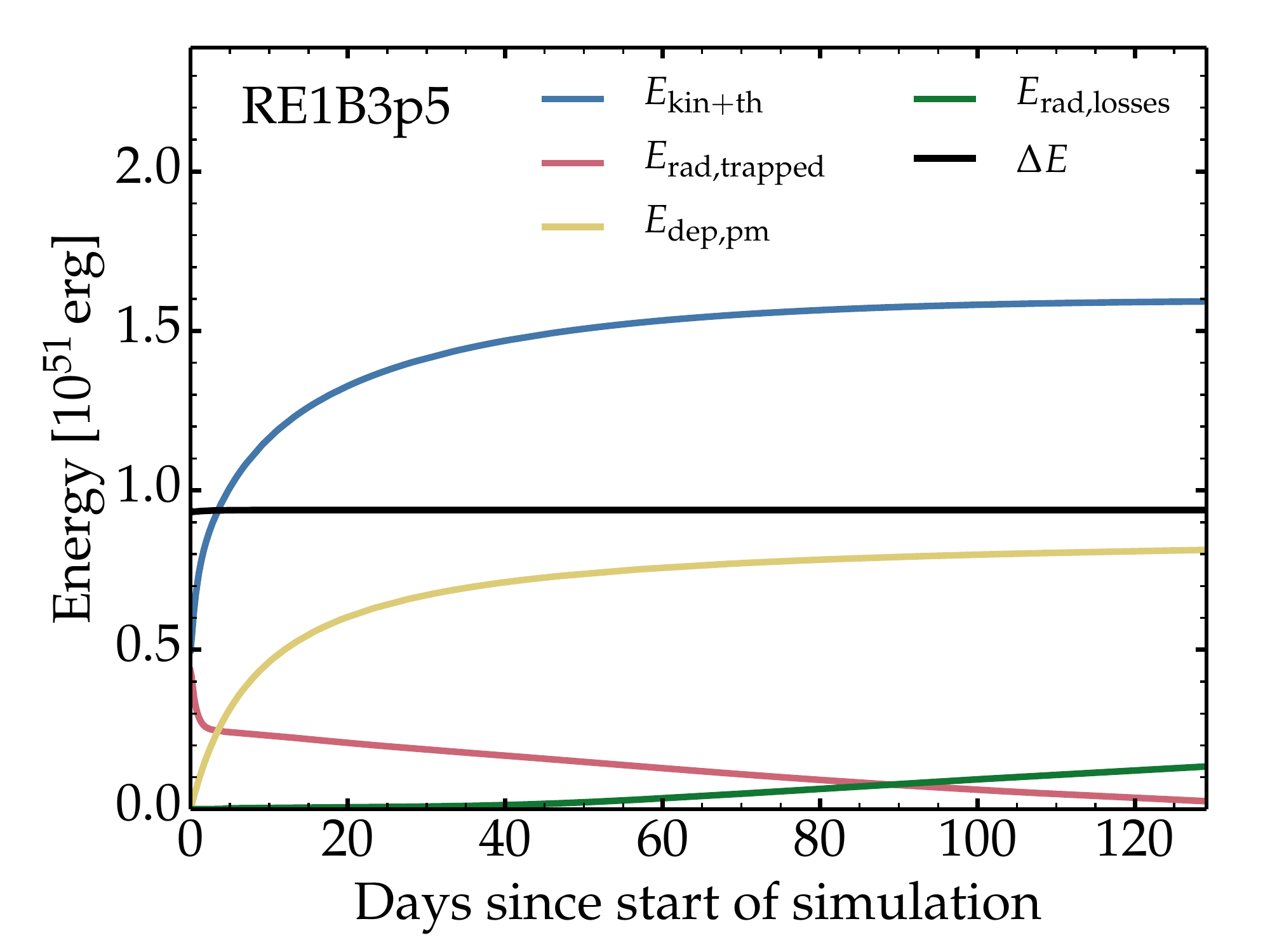}
\includegraphics[width=\hsize]{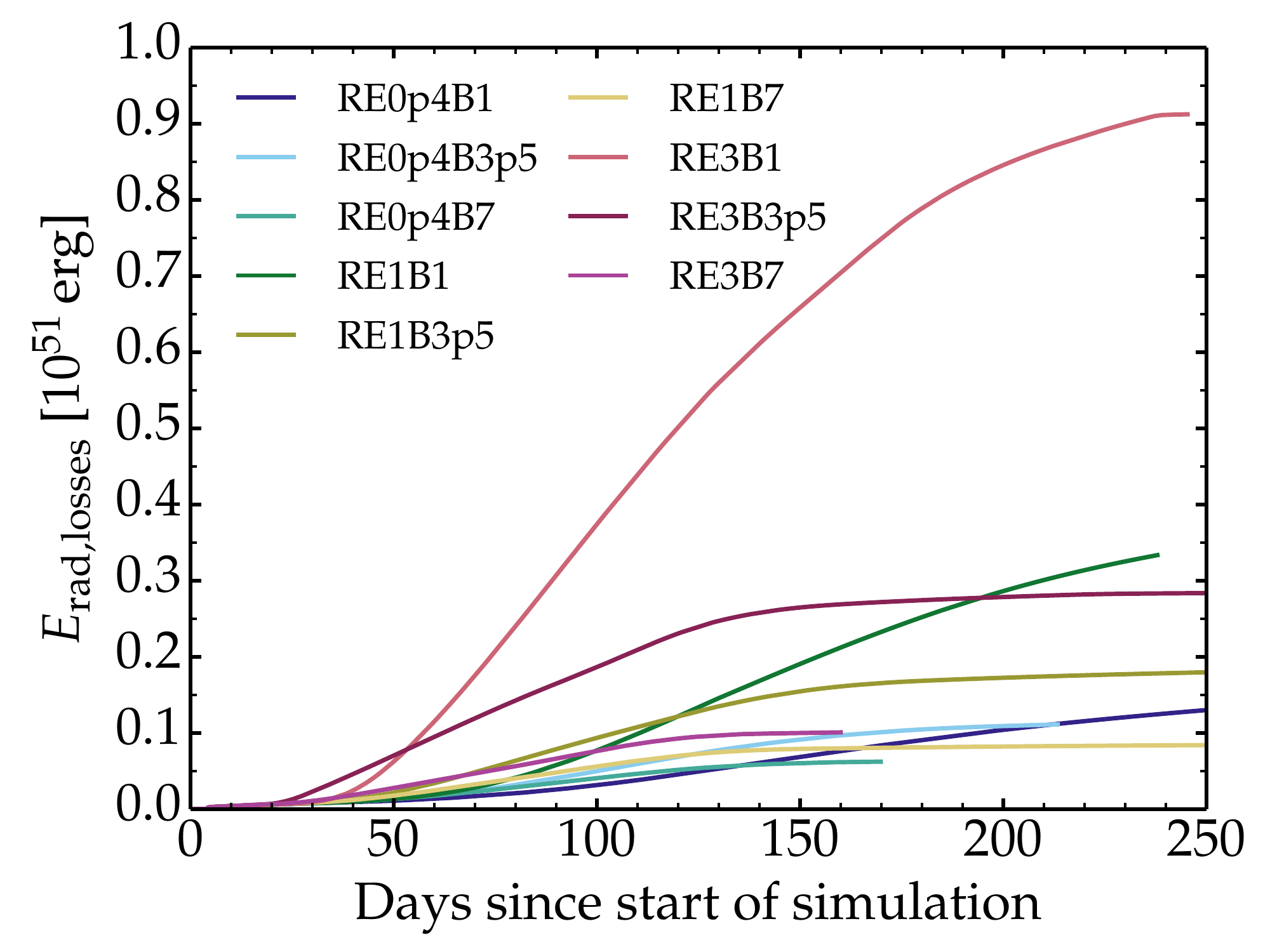}
\includegraphics[width=\hsize]{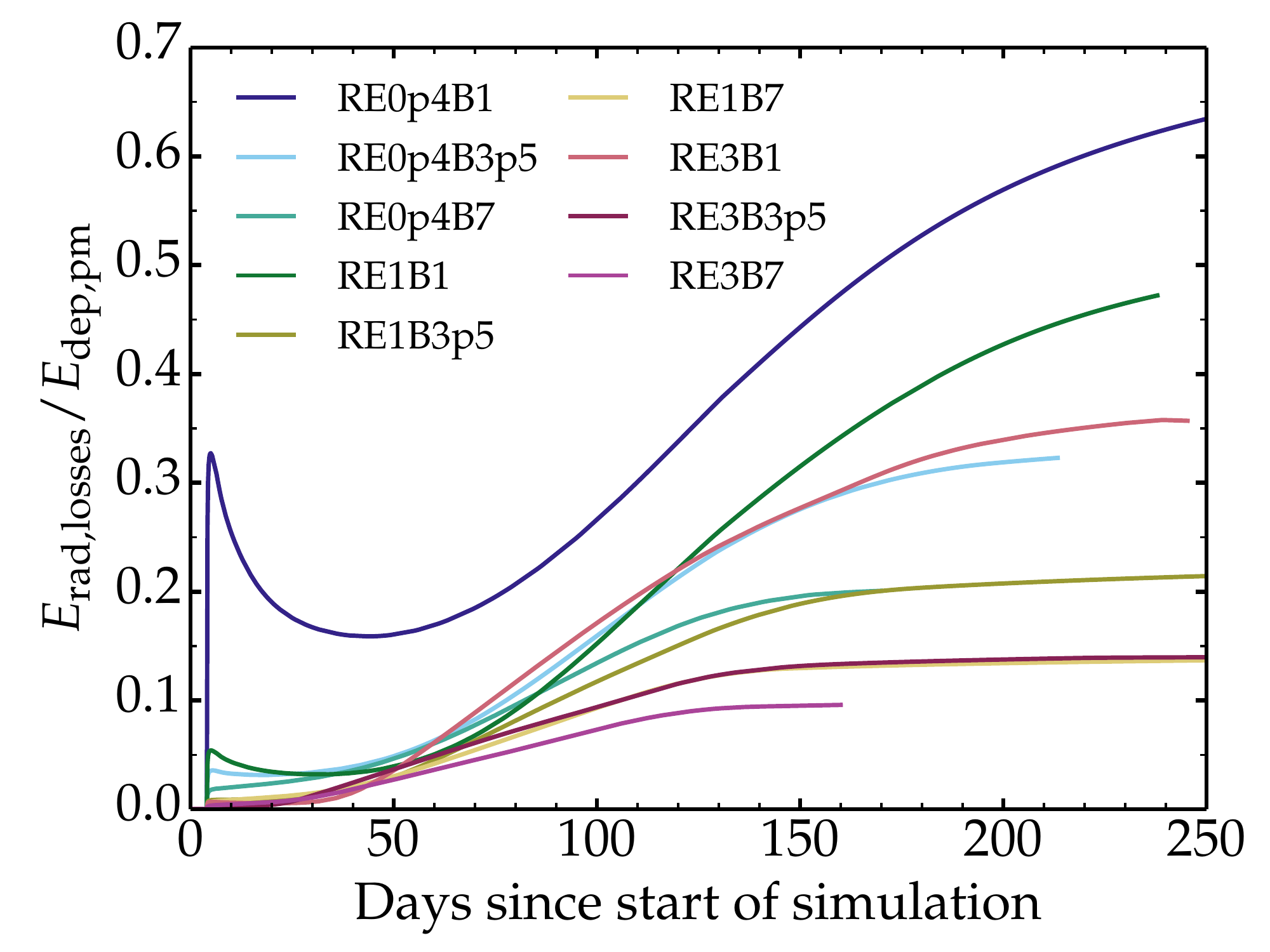}
\caption{Top:
Evolution of the various energy components in the
RE1B3p5 simulation. We show $E_{\rm kin+th}$, which
sums the kinetic and thermal (i.e., gas) energy;
$E_{\rm rad, trapped}$, which is the trapped radiation energy;
$E_{\rm dep, pm}$, which is the energy deposited by the magnetar;
$E_{\rm rad, losses}$, which is the radiative energy streaming out
through the outer grid boundary; and
$\Delta E \equiv E_{\rm kin+th} + E_{\rm rad, trapped} +
E_{\rm rad, losses} - E_{\rm dep, pm}$, which tracks conservation of energy.
The start of the simulation is 1.15\,d in all simulations --- any magnetar
energy deposited prior to that is ignored.
Middle and bottom: Evolution of the radiative losses (middle) and of
the fraction of the magnetar deposited energy that escapes as radiation
as a function of time (bottom) for the magnetar-powered SN models
based on the RSG progenitor.
Additional illustrations are provided in the appendix.
\label{fig_ener}
}
\end{figure}

\subsection{Energetics}

Figure~\ref{fig_ener} illustrates the energetics in our simulation RE1B3p5.
In the top panel, we show the evolution of the total gas energy $E_{\rm kin+th}$
(kinetic plus thermal contributions), the trapped radiation energy $E_{\rm rad, trapped}$,
the cumulative energy deposited by the magnetar $E_{\rm dep, pm}$,
and the cumulative radiative losses through
the outer boundary $E_{\rm rad, losses}$.
We also plot the quantity
$\Delta E \equiv E_{\rm kin+th} + E_{\rm rad, trapped} +
E_{\rm rad, losses} - E_{\rm dep, pm}$, which should be constant.
This illustration shows that initially (the model is within hours of shock breakout)
the trapped radiation energy is rapidly turned into kinetic energy while the
supply of energy from the magnetar boosts both the trapped radiation energy
and the ejecta kinetic energy. At 120\,d, a little more than 10$^{50}$\,erg has
been radiated to infinity, which is about ten times what this model would radiate
in the absence of a magnetar (assuming that a representative Type II SN, not
influenced by a central source,  radiates 10$^{49}$\,erg over its entire lifetime).
In this model RE1B3p5, the bulk of the magnetar energy
has been channeled into ejecta kinetic energy, but yielding only a 60\%
increase in ejecta kinetic energy (not the factor of ten above for the radiative energy losses)
over the value it would have had in the absence of a magnetar
(at 120\,d the magnetar in this model still has about $2 \times 10^{50}$\,erg to radiate).

The middle and bottom panels of Fig.~\ref{fig_ener} illustrate how efficiently
the magnetar energy is channelled into escaping radiation
(referred to as $E_{\rm rad, losses}$).
Nearly 10$^{51}$\,erg is radiated away in model RE3B1, while the rest
of the models falls in the range $0.05-0.3 \times 10^{51}$\,erg. The luminosity
boost over a standard SN II therefore goes from minor ($5 \times$)
to very large ($100 \times$).
When normalized to the current cumulative energy deposited by the magnetar,
the most efficient ``engines" are the magnetars with long spin down time scales,
that is, those with a lower magnetic field and/or a lower initial rotation energy (longer
periods). These trends and quantities would not be significantly altered if we had
started the magnetar energy deposition at the magnetar birth because
it is only radiation emitted on a long time scale that boosts the bolometric
luminosity of the SN \citep{KB10}.

\begin{figure*}
\includegraphics[width=0.49\hsize]{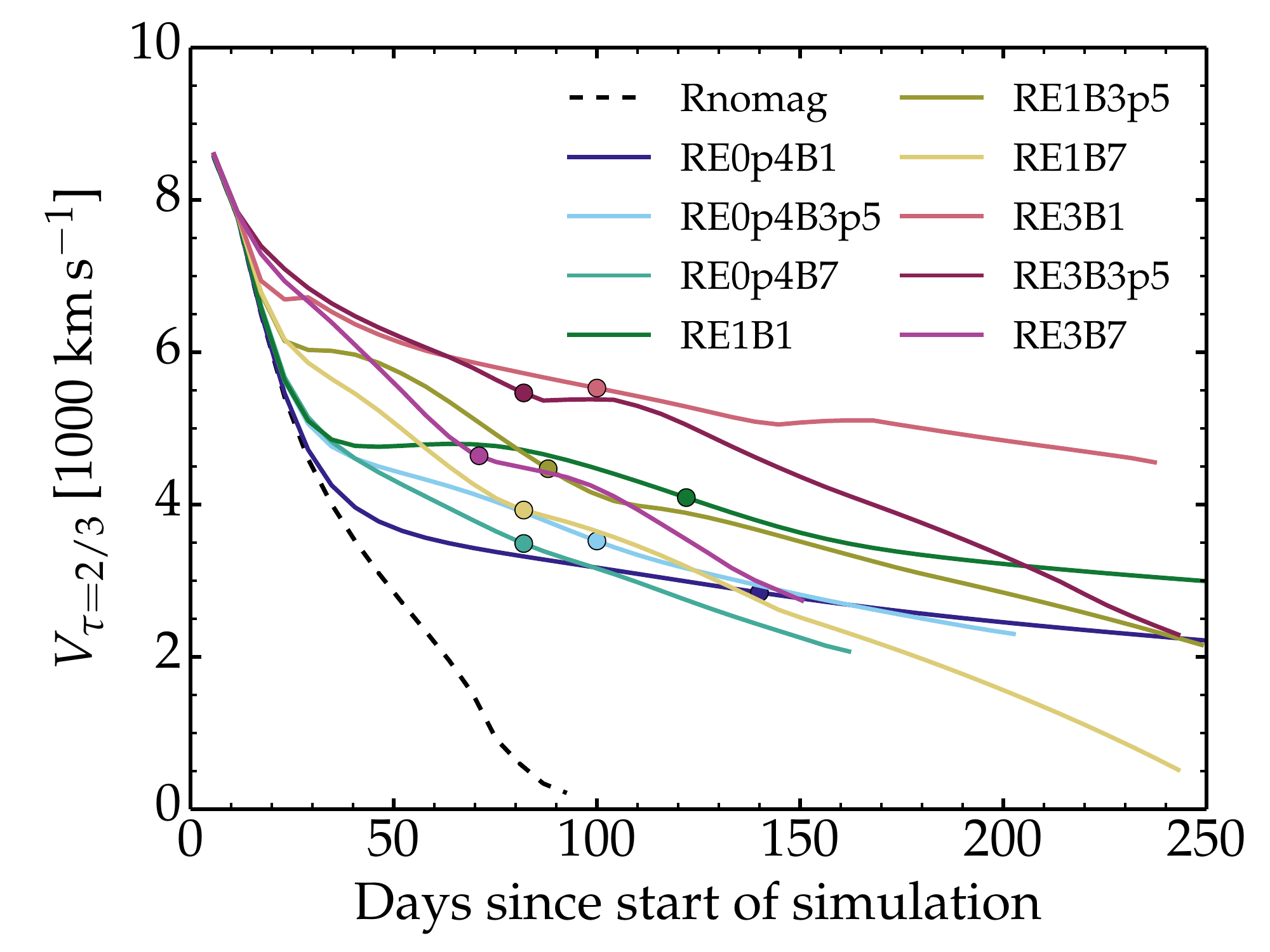}
\includegraphics[width=0.49\hsize]{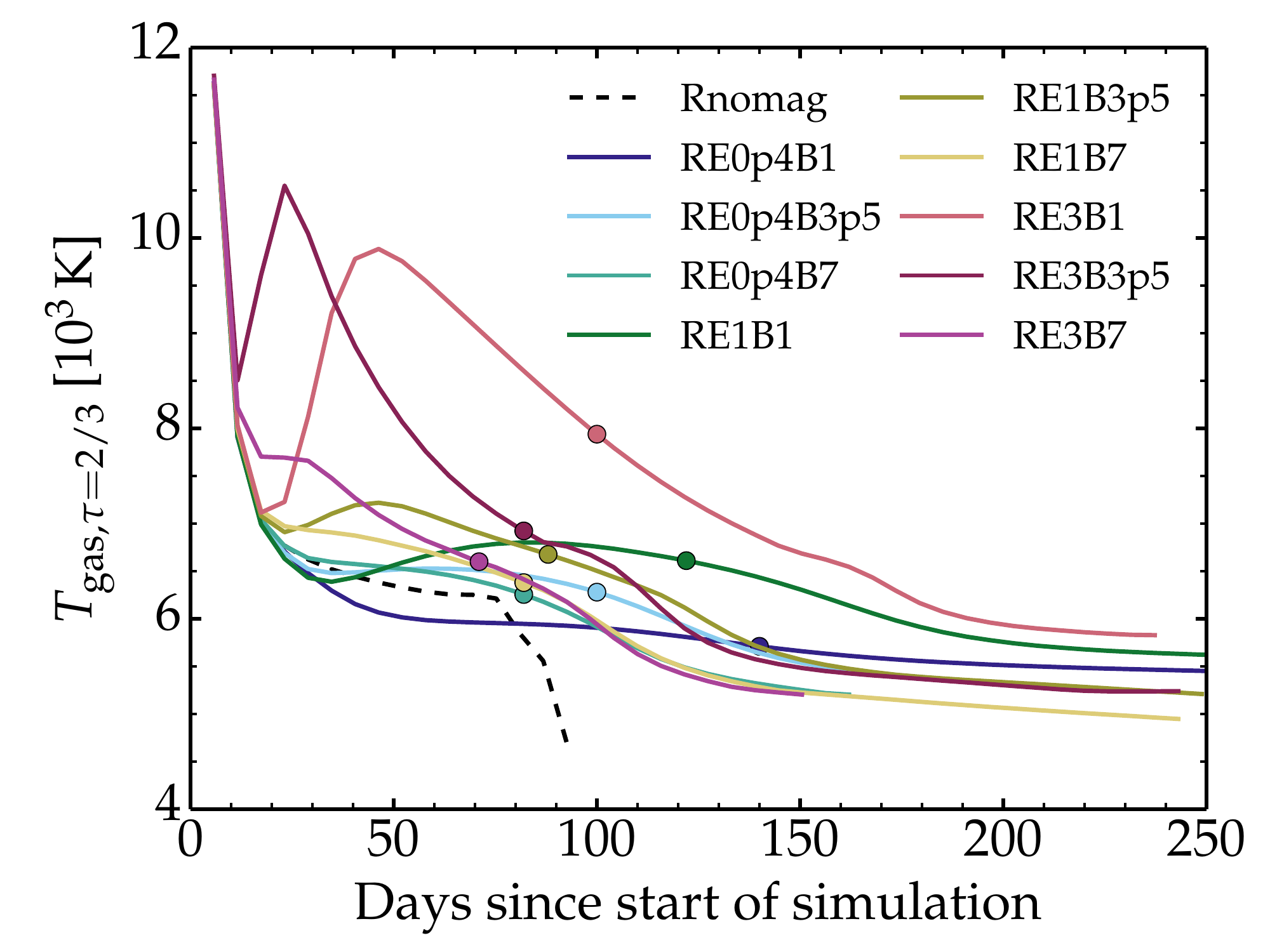}
\caption{
Evolution of the photospheric velocity with time in the \heracles\ simulations
for the RSG explosion model powered by a variety of magnetar energy and field strength.
The colored dot on each curve corresponds to the time at which we compute
the spectra shown in Fig.~\ref{fig_spec}.
}
\label{fig_phot}
\end{figure*}

\subsection{Photospheric properties}

We show the evolution of the photospheric properties for our magnetar-powered
SN models arising from the explosion of a RSG progenitor in Fig.~\ref{fig_phot}.
Compared to the model without magnetar (dashed line), the photospheric velocity
is at all times larger, and by a large factor after 100\,d since the no-magnetar model
is by then optically thin. The effect illustrated here does not arise from a change
in ejecta kinetic energy (which is essentially fixed after $50-100$\,d; see Fig.~\ref{fig_ener}
and appendix)
but stems instead from the higher ionization of the ejecta.
In the presence of a magnetar, the recombination is inhibited or delayed.
The near plateau in the photospheric velocity after $50-100$\,d implies that
the photosphere resides in a narrow range of ejecta mass shells
(until $>250$\,d in some models).

\begin{figure}
   \includegraphics[width=\hsize]{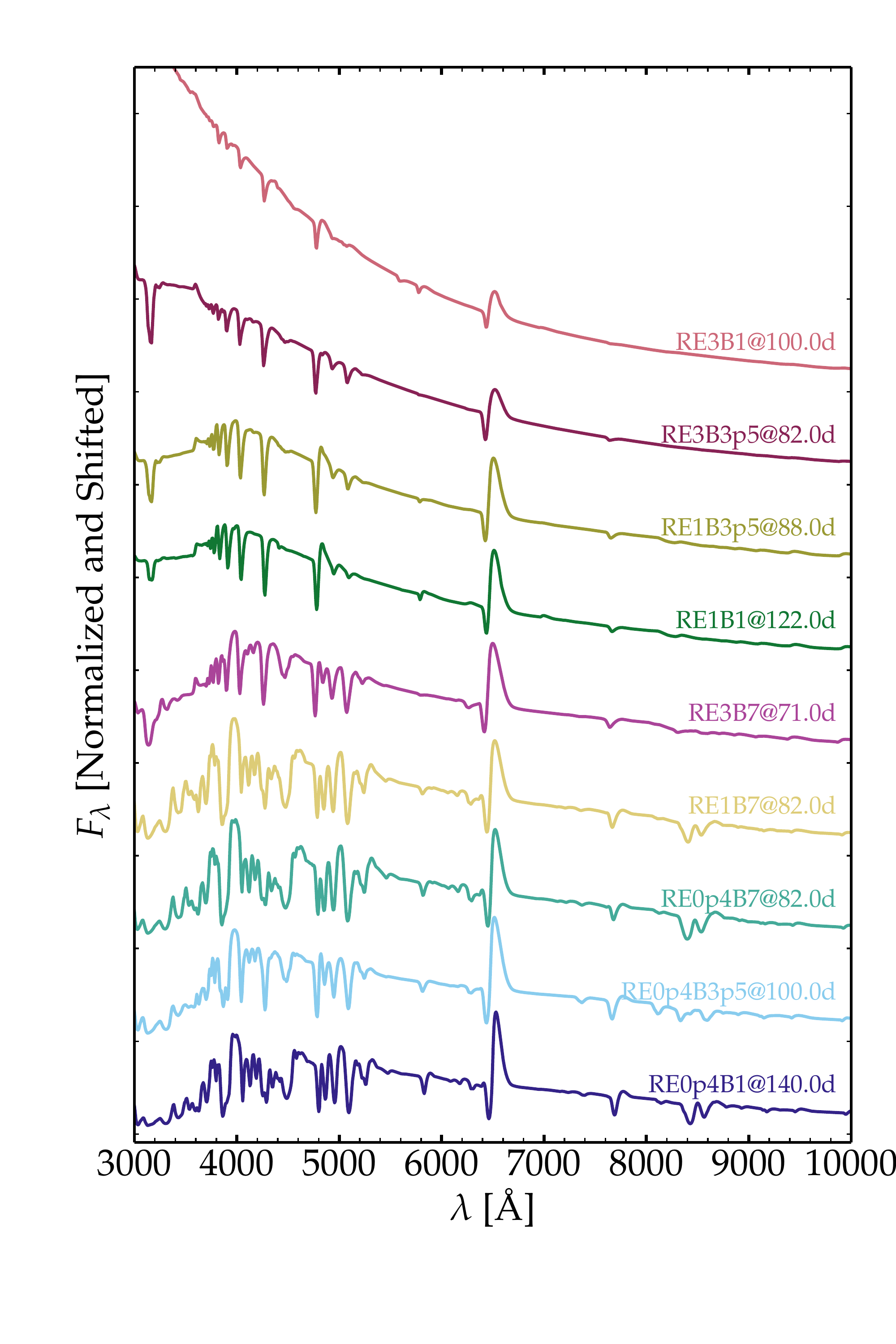}
\caption{Model spectra computed with \cmfgen\ for the magnetar-powered
RSG explosion models and corresponding to an epoch around bolometric
maximum (see label for model name and post-explosion epoch).
We stack the spectra vertically starting with the reddest at the
bottom (model Re0p4B1) and progressing upward toward bluer spectra.
Each flux spectrum is normalized at 10000\,\AA\ and shifted
vertically for visibility (the ordinate tick mark gives the zero-flux level).
\label{fig_spec}
}
\end{figure}

The photospheric temperature is larger at all times in the presence of a
magnetar.\footnote{This holds except for model RE0p4B1, which is slightly cooler at the
photosphere. The energy released by the magnetar does not in principle need to produce
a higher  photospheric temperature compared to the same model without a magnetar.
Indeed, the influence of the magnetar may simply be
to push the photosphere outward in mass/velocity space, while the photospheric temperature
remains close the recombination temperature of hydrogen. In a SN II, the photosphere
at the recombination epoch coincides with the layer where hydrogen changes ionization,
from neutral to ionized. As long as there is such a recombination front, the photosphere
is at roughly the same temperature, around 6000\,K.}
In many models, this temperature is close to the recombination temperature of hydrogen,
which implies that the boost to the model luminosity is caused by the greater photospheric radius.

Interestingly, in the models with $E_{\rm pm}>10^{51}$\,erg, the initial decline of the
photospheric temperature is halted at $10-40$\,d after the magnetar birth and the temperature
starts rising again. In the models with the faster rotating magnetars (RE3B1 and RE3B3p5),
the photospheric temperature reaches a maximum of about 10,000\,K, which is sufficient
to re-ionize hydrogen (in model RE3B7, the spin-down time scale is so short that most
of the energy goes into work to accelerate the ejecta and the boost to $T_{\rm phot}$ is
smaller than for models  RE3B1 and RE3B3p5).

A non-monotonic evolution of the photospheric temperature
and of the location at maximum absorption in
He\one\,5875\,\AA\ have been observed in the Type Ib SN\,2005bf \citep{folatelli_05bf_06}.
This is consistent with the magnetar influence invoked to reproduce the light curve
of SN\,2005bf \citep{maeda_05bf_07}.

\begin{figure*}
   \includegraphics[width=\hsize]{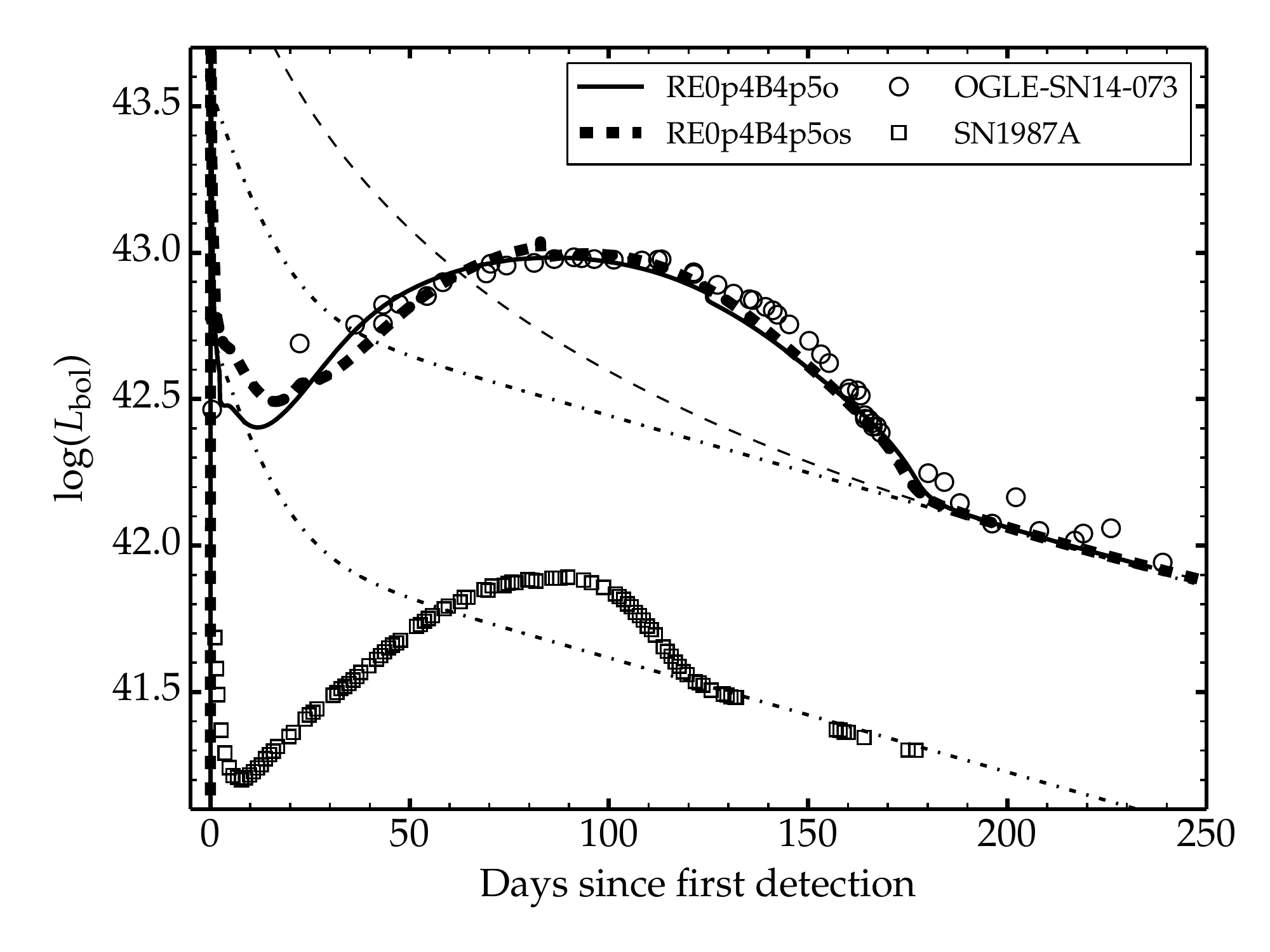}
\caption{Comparison of the bolometric light curves of magnetar-powered
models RE0p4B4p5o  (thick dashed line) and RE0p4B4p5os (thin line) with
that inferred for  OGLE-SN14-073 (empty circles; \citealt{terreran_slsn2_17}) and SN\,1987A
(empty squares; \citealt{hamuy_87A_88}). The origin of the  $x-$axis is the time of the first
signal detection at the outer grid boundary for the models ($\sim$\,4\,d after
the start of the simulations), or the inferred time of explosion for the observations.
We add the magnetar power of these two models (thin dashed line) and the radioactive
decay power from 0.07\,\msun\ of \nifs\ (to match the nebular luminosity of SN\,1987A;
dash-dotted line). To power the nebular luminosity of OGLE-SN14-073 with radioactive
decay requires 0.47\,\msun\ of \nifs\ (upper dash-dotted line).
\label{fig_ogle}
}
\end{figure*}

\subsection{Spectral properties at bolometric maximum}

We conclude this section by discussing the optical spectra for our grid
of models based on the RSG progenitor (Fig.~\ref{fig_spec}). This
post-treatment of non-monotonic ejecta with \cmfgen\ is currently
only possible during the photospheric
phase so we choose the unambiguous time of bolometric maximum
for this illustration.

The maximum light spectra shown in Fig.~\ref{fig_spec} cover
a range of optical colors, reflecting the range in photospheric temperature
(Fig.~\ref{fig_phot}). Bluer spectra appear like standard Type II SNe
at early times, when the conditions at the photosphere are ionized
or partially ionized. In model RE3B1 at 100\,d, we recognize
lines of H\one, He\one, and N\two, as in the earliest spectra
for SN\,1999em \citep{DH05a,DH06}. Redder spectra appear
like the standard Type II SNe  at later times in the photospheric phase,
when the conditions at the photosphere are partially neutral (and an
H recombination front has formed; \citealt{DH11_2p}).

The line widths are, however, anomalously low for the SN luminosity.
Indeed, the photospheric velocities at bolometric maximum are
about 4000\,\kms\ (Fig.~\ref{fig_phot} and Table~\ref{tab_sum}),
which is standard for a Type II \citep{hamuy_03},
but the luminosity is up to $10-50$ times larger than standard.
In our choice of magnetar properties, the energy released by the magnetar
boosts primarily the radiation budget and affects little the kinetic energy
(choosing a magnetar with a larger rotational energy and larger magnetic field
would do the opposite).
In doing so, it breaks the tight correlation between brightness and expansion
rate inferred from the observation of Type II SNe \citep{hamuy_03}.
Some observed Type II SNe, with no apparent sign of interaction,
appear over-luminous for their expansion rate,
for example, LSQ13fn \citep{polshaw_lsq13fn_16} or SN\,2006V \citep{taddia_2pec_12},
and this may be a signature of a magnetar influence.


\section{Comparison to observations}
\label{sect_obs}

Few observations of super-luminous Type II SNe suggest a magnetar
influence. \citet{KB10} propose a magnetar with $B_{\rm pm} = 2 \times 10^{14}$\,G,
$E_{\rm pm} = 5 \times 10^{51}$\,erg (rotation period of 2\,ms), and a H-rich ejecta
of 5\,\msun\ to explain the light curve of SN\,2008es \citep{gezari_08es_09,miller_08es_09}.
Such a low mass ejecta is hard to fit within the context of an H-rich massive
star progenitor so we leave this object aside for now.

Here, we choose a case that seems to be less ambiguous.
We compare our simulations to the super-luminous Type II SN
named OGLE-SN14-073 \citep{terreran_slsn2_17}, which
exhibits a bolometric light curve similar to that of SN\,1987A but
brighter by a factor of ten and significantly broader (Fig.~\ref{fig_ogle},
big black circles) --- the time integrated luminosity up to $>200$\,d is about 10$^{50}$\,erg.
The photospheric velocity inferred during the high brightness
phase is about 6000\,\kms. \citet{terreran_slsn2_17} propose
a RSG progenitor with an explosion model yielding a very large ejecta
mass of about 60\,\msun\ and kinetic energy of $12 \times 10^{51}$\,erg.
Their quoted uncertainties are however quite large.

  We have compared our grid of models and found that RE0p4B3p5 and RE0p4B7
  match closely the bolometric light curve. These models
arise from an ejecta of 11.9\,\msun\ and a kinetic energy of about $10^{51}$\,erg,
so 10 times less energetic and 5 times less massive than proposed for OGLE-SN14-073 .
Because their luminosity is too large at late times and too small at early times,
we performed a new simulation (using the same initial ejecta model) with a higher magnetar field
($B_{\rm pm} = 4.5 \times 10^{14}$\,G) and a broader energy deposition profile
(in order to hasten the magnetar influence and produce a higher luminosity at early times ---
the alternative of a greater progenitor radius would also yield a greater brightness at
early times).

In Fig.~\ref{fig_ogle}, we compare the bolometric light curve of this
new model RE0p4B4p5o with the observations of  OGLE-SN14-073, and include SN\,1987A for reference.
Model RE0p4B4p5o
matches quite closely OGLE-SN14-073. However, the photospheric velocity of these
three models is about 4000\,\kms\ at maximum, which is 50\% lower than
inferred for OGLE-SN14-073. Scaling the initial density and velocity by 50\%,
as well as scaling the temperature by 22\% (the model is pre-breakout hence
half the total shock deposited energy is radiation, the other half is kinetic)
we produce a new model RE0p4B4p5os (total mass of 17.8\,\msun\
and a kinetic energy of $2.67 \times 10^{51}$\,erg).
This choice of scaling is to increase the velocity while maintaining the
same diffusion time (which scales as $\sqrt{M/V}$).
RE0p4B4p5os yields a similar light curve to OGLE-SN14-073.
However, the photospheric velocity at bolometric maximum is only increased to 4280\,\kms,
hence still significantly lower than the inferred value of 6000\,\kms\ for OGLE-SN14-073.
Perhaps the kinetic energy is not underestimated and there is instead an error in inferring
the photospheric velocity. The magnetar influence
may affect line profiles
such that the location of maximum line absorption occurs at a Doppler velocity
significantly greater than the photospheric velocity value we report (which accounts
only for electron scattering).
Time dependence might also affect line profiles by preventing the formation of a steep recombination
front as normally seen in standard Type II SNe \citep{UC05,D08_time}.
Numerical explorations of magnetar-powered Type II SNe with \cmfgen\ show that
spectral lines can remain very broad until late times, and broader than
predicted using steady state (Dessart, in preparation).
This aspect requires further study. But it seems likely that OGLE-SN14-073 is a unique Type II event,
combining a massive and very energetic ejecta with a magnetar having
$B_{\rm pm} = 4.5 \times 10^{14}$\,G and $E_{\rm pm} = 0.4 \times 10^{51}$\,erg.

   The ejecta corresponding to RE0p4B4p5os has a kinetic energy and mass greater
than standard for a Type II SN but less extreme than proposed by
 \citet{terreran_slsn2_17}. Although not explicitly stated in their paper,
 the model of Terreran et al. requires 0.47\,\msun\ of \nifs, while in our
 magnetar powered model we assume no \nifs. The influence on the
 bolometric light should be similar since the power released in each process
 is similar (see upper dash-dotted and dashed lines in Fig.~\ref{fig_ogle}).
 In the radioactive decay scenario, this \nifs\ mass makes sense since the
 total decay energy from 1\,\msun\ of \nifs\ is $\sim 2 \times 10^{50}$\,erg,
and the time integrated luminosity of OGLE-SN14-073 is about 10$^{50}$\,erg.
The difference between a magnetar-powered model and a \nifs-powered model
is however non trivial when considering spectra and colors \citep{d12_magnetar}.
In OGLE-SN14-073, the presence of a Type II spectrum at all times
with little evidence for metal lines \citep{terreran_slsn2_17} is hard to combine with
the presence of a large progenitor CO core required to yield a large \nifs\ mass.
In explosion models powered by \nifs, the spectrum formation region at
the time of bolometric maximum is located in metal rich regions \citep{d13_pisn},
which is in conflict with the observations of OGLE-SN14-073.
If the power source is a magnetar, the ejecta composition may instead be dominated by H and He,
with is compatible with the Type II spectrum observed at all times in OGLE-SN14-073.

At the time of bolometric maximum, the luminosity is a factor of 2 above the concomitant
magnetar power in models RE0p4B4p5o and RE0p4B4p5os  --
similar offsets are present in the set of simulations show in Fig.~\ref{fig_pm_rsg}.
An 80\% offset is seen for SN\,1987A between the bolometric maximum and the
concomitant decay power from 0.07\,\msun\ of \nifs. Arnett rule \citep{arnett_82}
states that the peak bolometric luminosity should be equal to the power from
which it derives (magnetar radiation, radioactive decay).


\section{Conclusion}
\label{sect_conc}

   We have presented 1-D gray radiation-hydrodynamics simulations of
 magnetar-powered Type II SNe performed with the code \heracles\
 \citep{gonzalez_heracles_07,vaytet_mg_11}. Our simulations are based on BSG
 and RSG progenitors, which reproduce roughly the properties
 of SN\,1987A \citep{DH10} and SN\,1999em \citep{d13_sn2p} if \nifs\ decay
 power (rather than magnetar power) is accounted for .
 We adopt a power source from a magnetic dipole for convenience. In practice,
 the power radiated by such a compact remnant may not follow such a
 smooth evolution.

    In our supergiant progenitors, the composition is dominated by H and He --
the contribution in metals from the He core is secondary.
In our magnetar-powered SN simulations, allowing for chemical stratification
using five representative species (H, He, O, Si,
and \nifs) makes little difference compared to assuming a composition typical
of a RSG H-rich envelope. Nearly all our simulations therefore assume a uniform
H-rich composition. Furthermore, we mimic multi-D fluid instabilities seen
in the simulations of \citet{chen_pm_2d_16,suzuki_pm_2d_17} by depositing
the magnetar energy of a range of masses, rather than over a narrow mass range
at the base. With this approach, the density and temperature structures
remain smooth at all times,  the light curves rise earlier and do not show
a late time bump in the photospheric phase.

   We find that early-time observations may help constrain the size of the progenitor
star. Provided the magnetar influence does not start too soon, the SN luminosity
arising from a BSG explosion is markedly lower than that from a RSG explosion.
By the time of maximum, the influence of the progenitor radius on expansion cooling
is swamped by the large energy release from the magnetar. The light curves from
the BSG and RSG magnetar-powered SNe overlap at and beyond maximum.
Similarly, the optical spectra at maximum are similar and one would not be able
to discern between the two progenitors from such a spectral information.

   We then present a grid of models for RSG explosions influenced by a magnetar.
For our set of magnetar field ($1.0-7.0 \times 10^{14}$\,G) and magnetar initial rotational
energy ($0.4-3.0 \times 10^{51}$\,erg), the resulting bolometric light curves
reach a broad peak at $50-150$\,d after explosion with a power of
$0.7-7.8 \times 10^{43}$\,\ergs.
The fraction of the magnetar energy channelled into (UV-optical-infrared)
SN radiation is higher for weaker field but in all cases the boost to the SN luminosity
is significant (between a factor of 5 and 100 compared to a standard Type II SN).
The magnetar influence delays the recombination of the ejecta and maintains
the photosphere at a larger radius (in a mass shell located further out in the ejecta).
In some cases, the magnetar energy deposition reverses the cooling and causes
the photosphere to heat up after a few months. SN\,2005bf exhibits a similar
phenomenon \citep{folatelli_05bf_06}, which is compatible with a magnetar origin
\citep{maeda_05bf_07}.
The maximum light spectra are analogous to those of standard SNe II, with a range
in colors that reflects the scatter in photospheric temperature (itself dependent on
the heating efficiency of the magnetar). However, the modest expansion rate and
huge brightness of these magnetar powered SNe breaks the brightness/expansion-rate
correlation observed in standard SNe II \citep{hamuy_03}.

   Amongst super-luminous SNe II, it appears that OGLE-SN14-073
   \citep{terreran_slsn2_17} may be interpreted
   as a magnetar powered SN. We find that our RSG model influenced by a
   magnetar with $E_{\rm pm} = 0.4 \times 10^{51}$\,erg and
   $E_{\rm pm} = 4.5 \times 10^{14}$\,G yields a satisfactory match to the bolometric
   light curve, although our model underestimates the inferred expansion rate.

  The main uncertainties in our simulations are the very approximate handling of the impact
of multi-dimensional fluid instabilities (which will introduce
clumping; \citealt{jerkstrand_slsnic_17}), the neglect of time
dependence in the non-LTE solution, and the neglect of non-thermal effects
associated with the energy injection from the magnetar.
   Further modeling is therefore needed, for example to test whether the line profile widths
   are broadened by the magnetar influence and time dependent effects,
   which could potentially lead to overestimating the ejecta kinetic energy and mass.

\begin{acknowledgements}

    We thank Giacomo Terrreran for providing the estimated
bolometric luminosity of OGLE-SN14-073.
LD thanks ESO-Vitacura for their hospitality.
This work utilized computing resources of the mesocentre SIGAMM,
hosted by the Observatoire de la C\^ote d'Azur, Nice, France.
This research was supported by the Munich Institute for Astro-
and Particle Physics (MIAPP) of the DFG cluster of excellence
``Origin and Structure of the Universe".

\end{acknowledgements}

\appendix

\section{Additional illustrations}
\begin{figure*}
\vspace{1cm}
   \includegraphics[width=0.33\hsize]{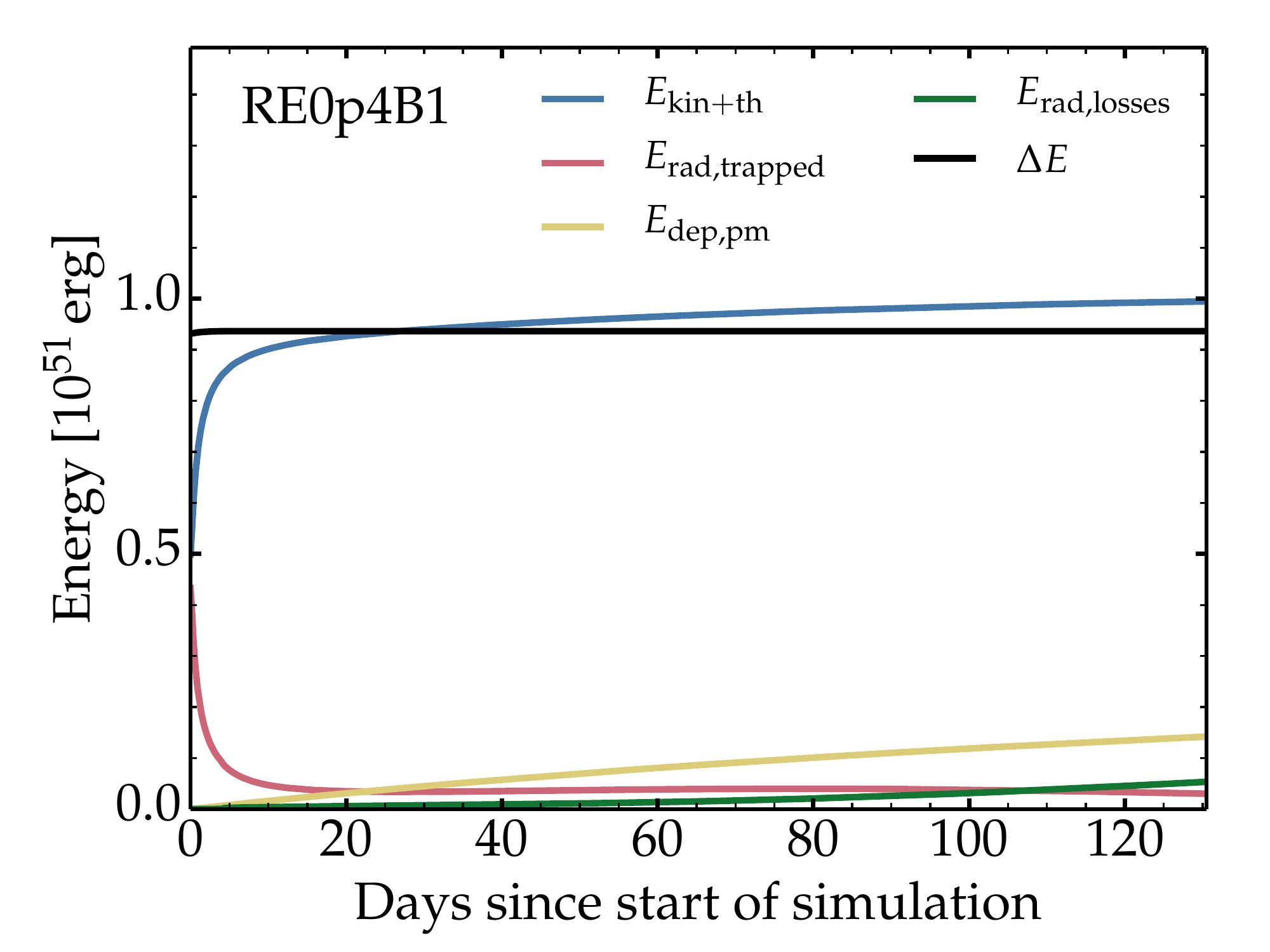}
   \includegraphics[width=0.33\hsize]{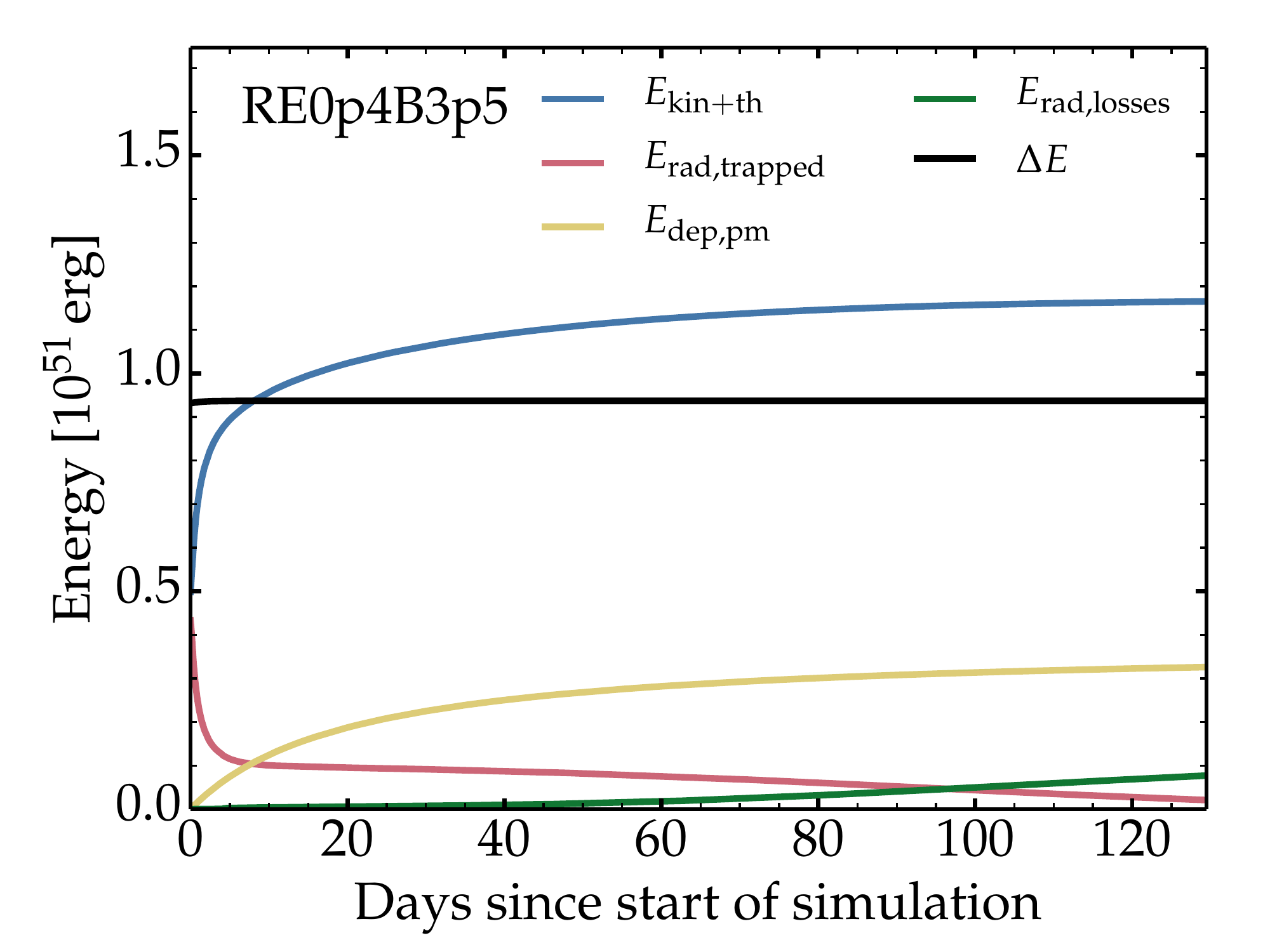}
   \includegraphics[width=0.33\hsize]{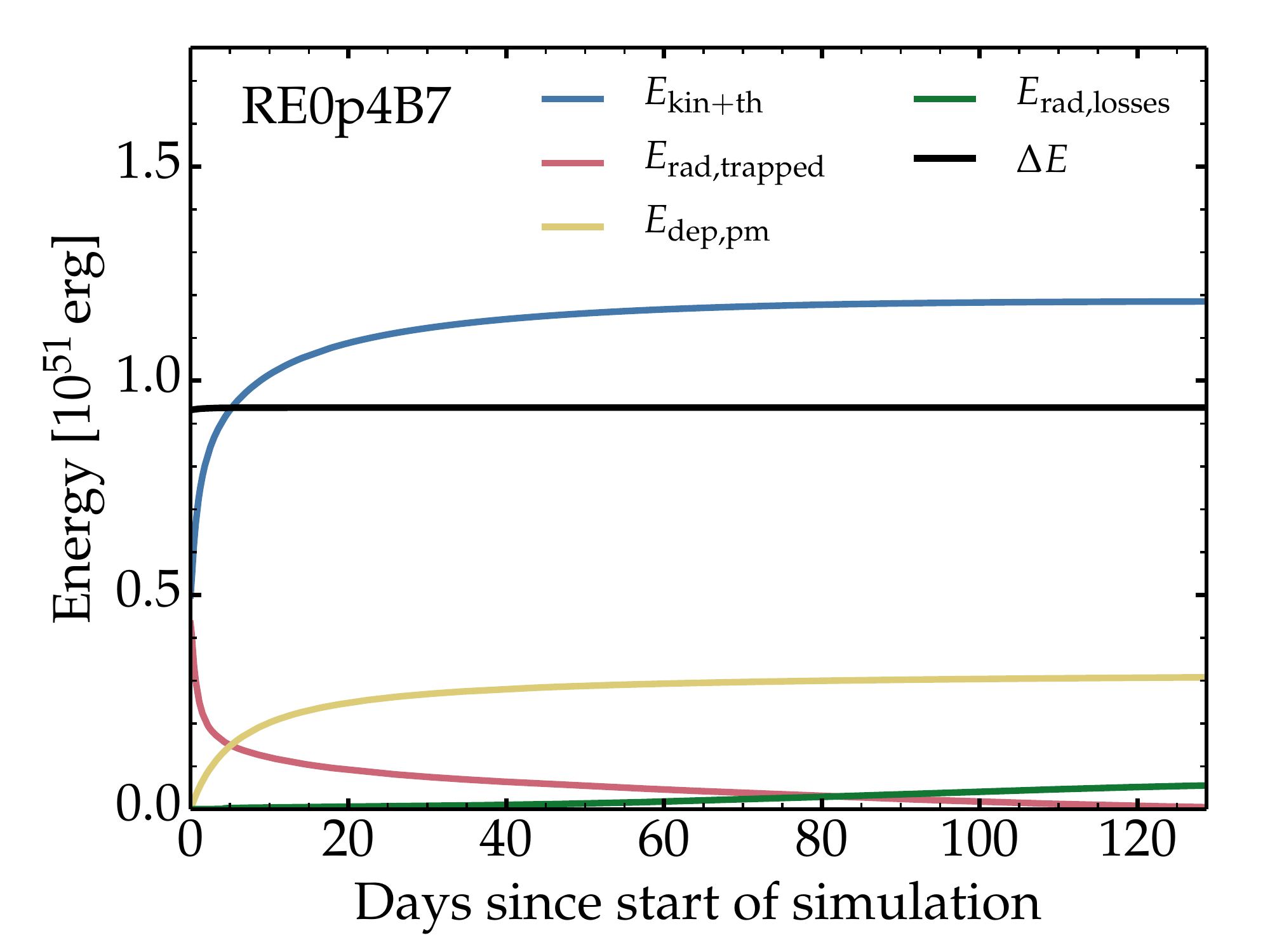}
   \includegraphics[width=0.33\hsize]{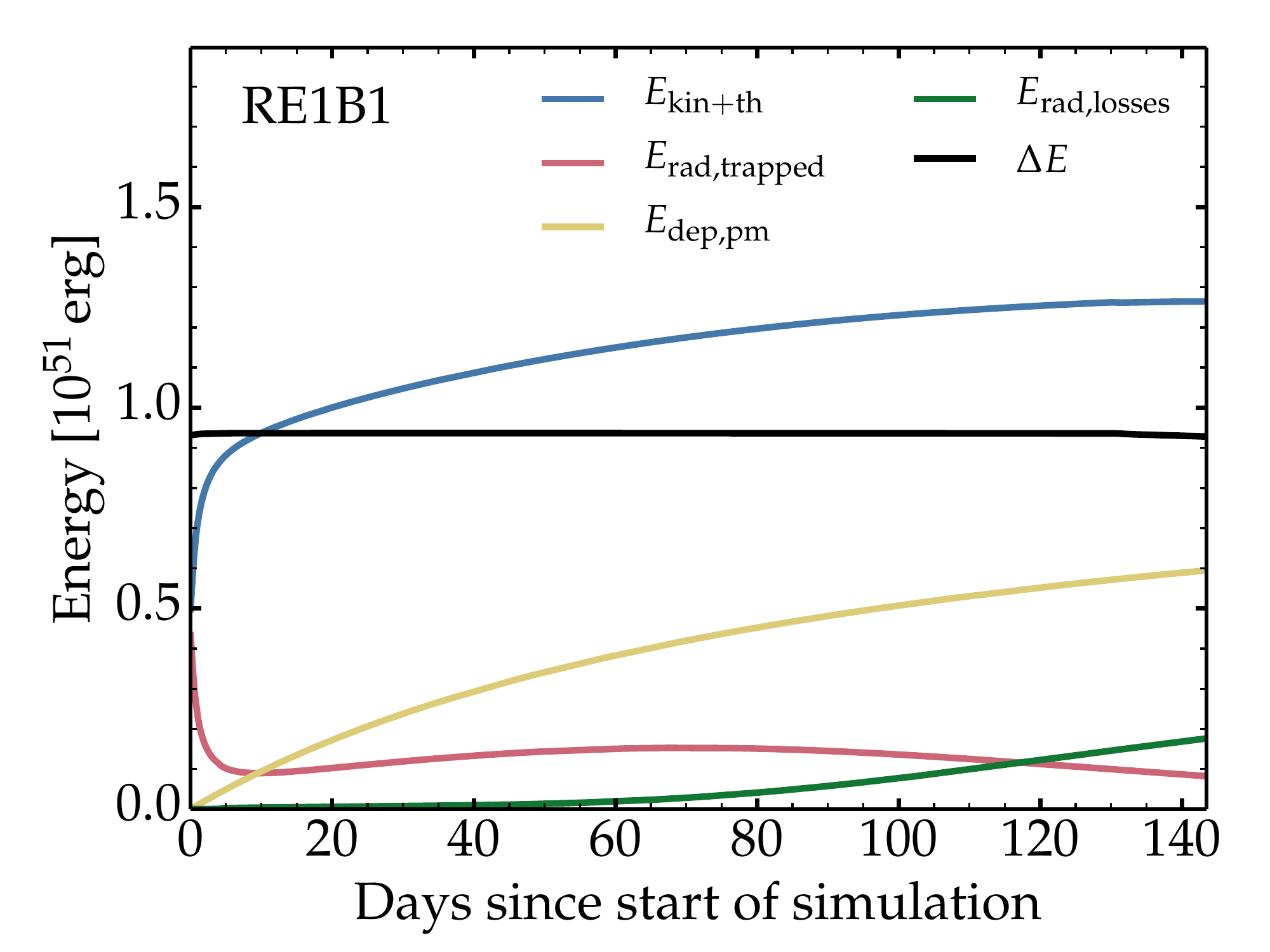}
   \includegraphics[width=0.33\hsize]{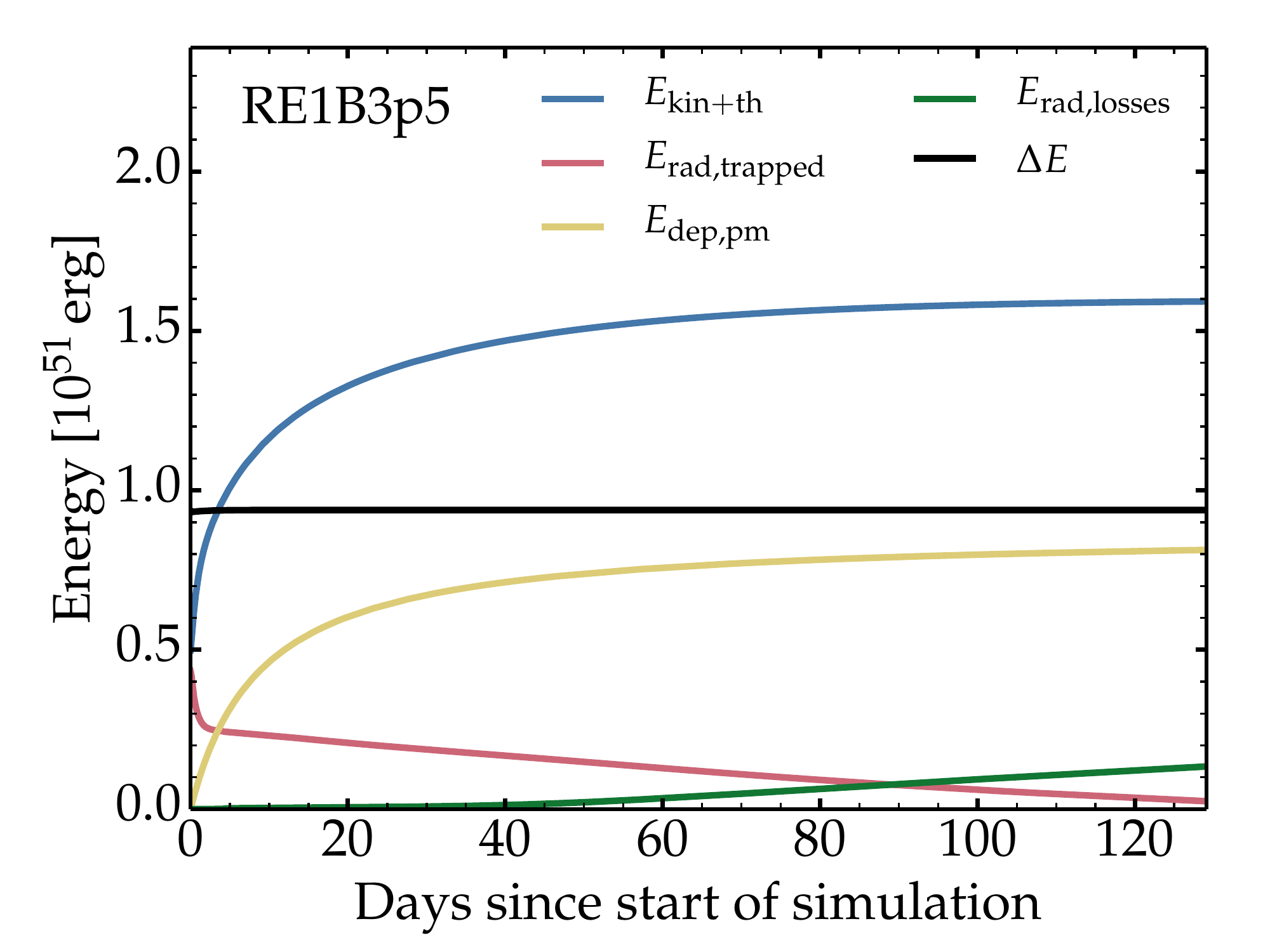}
   \includegraphics[width=0.33\hsize]{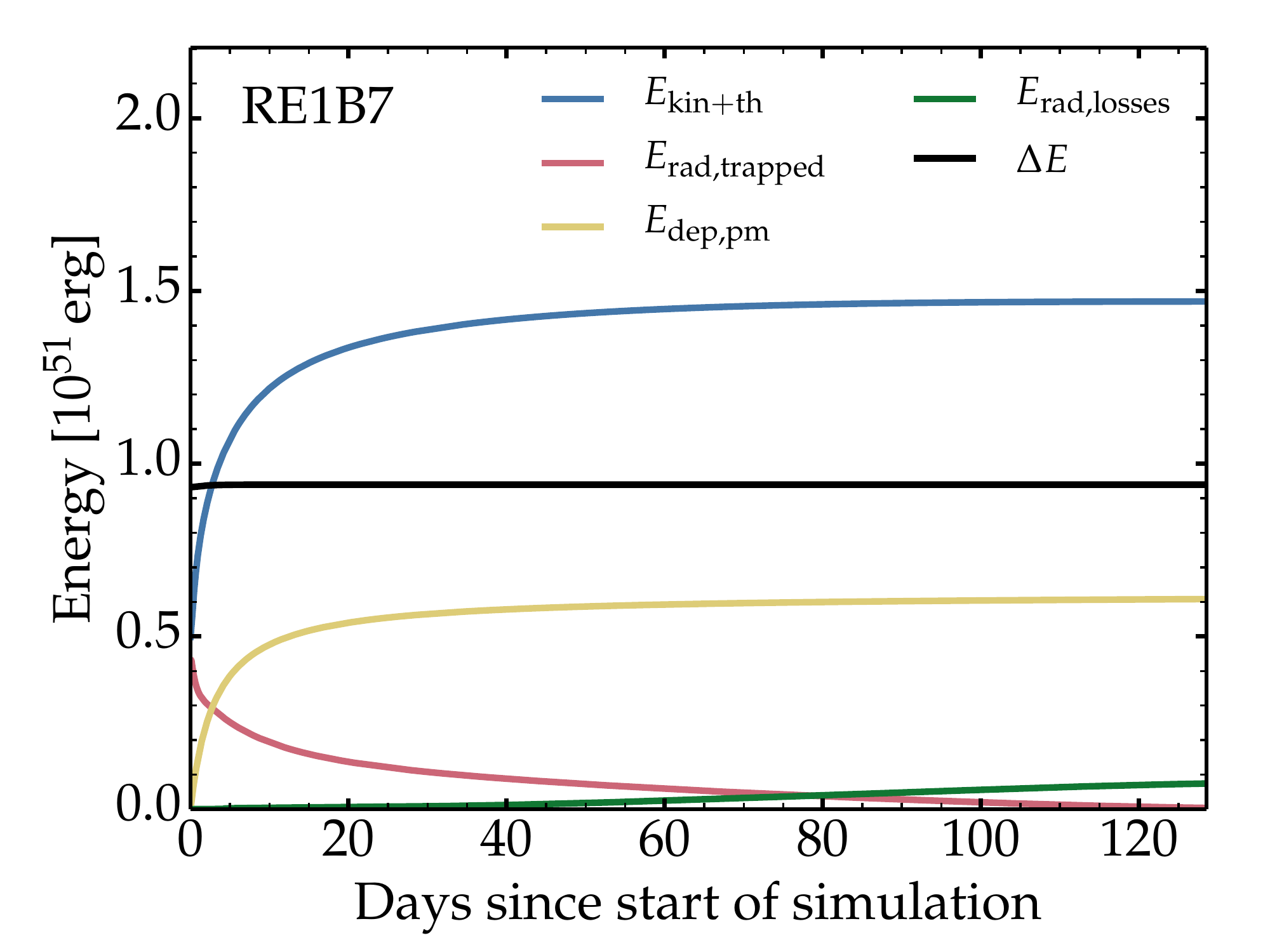}
   \includegraphics[width=0.33\hsize]{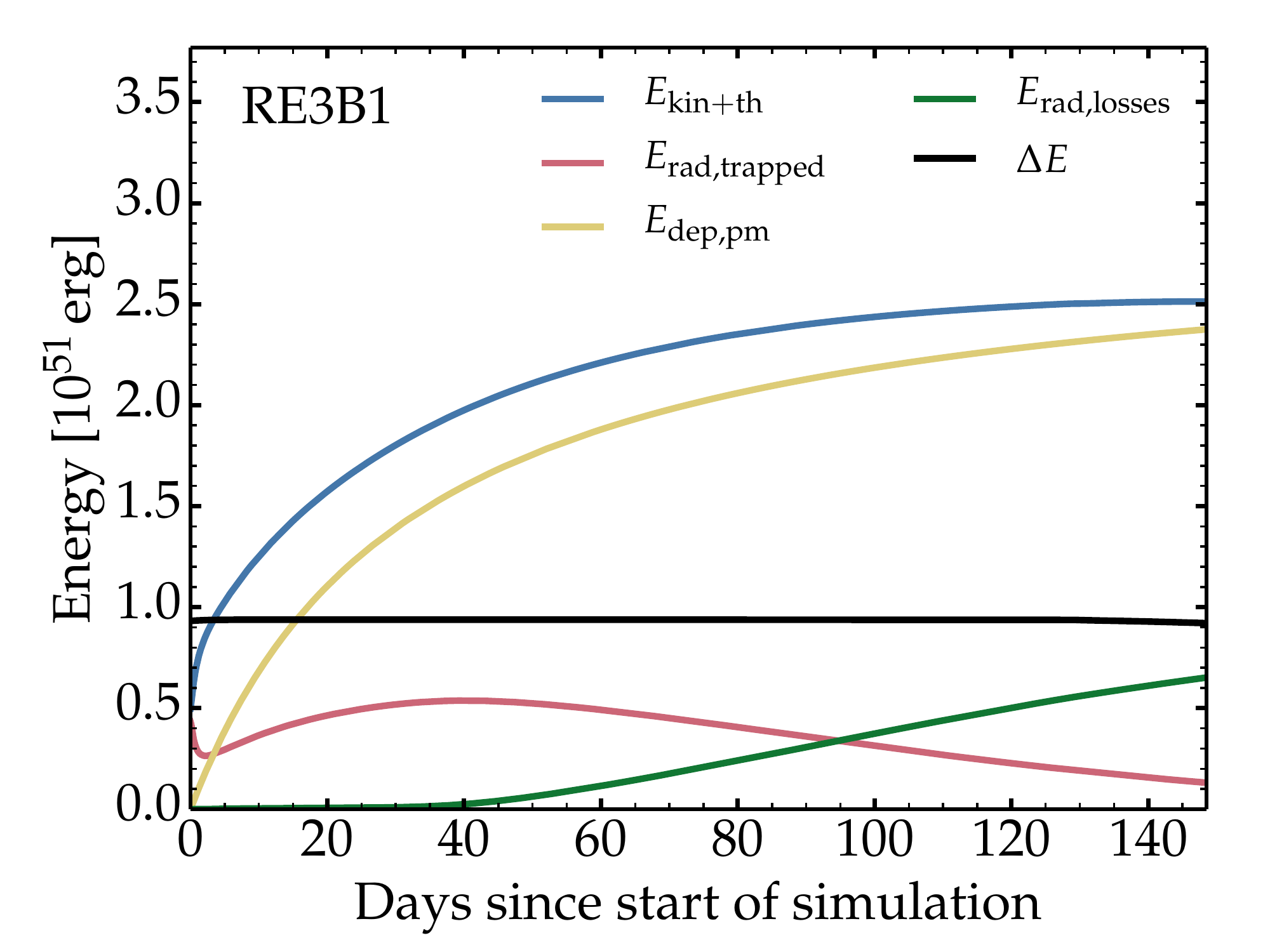}
   \includegraphics[width=0.33\hsize]{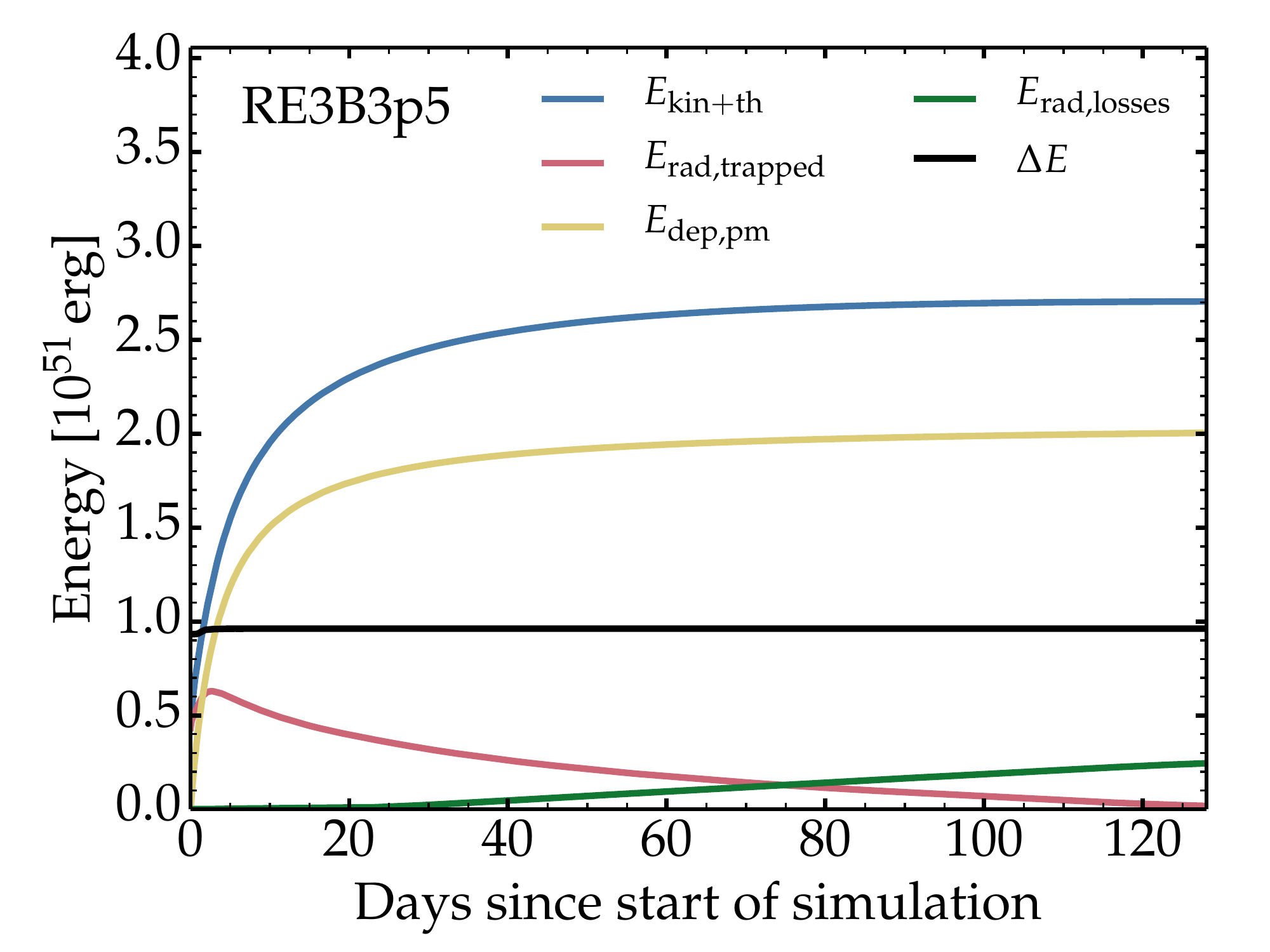}
   \includegraphics[width=0.33\hsize]{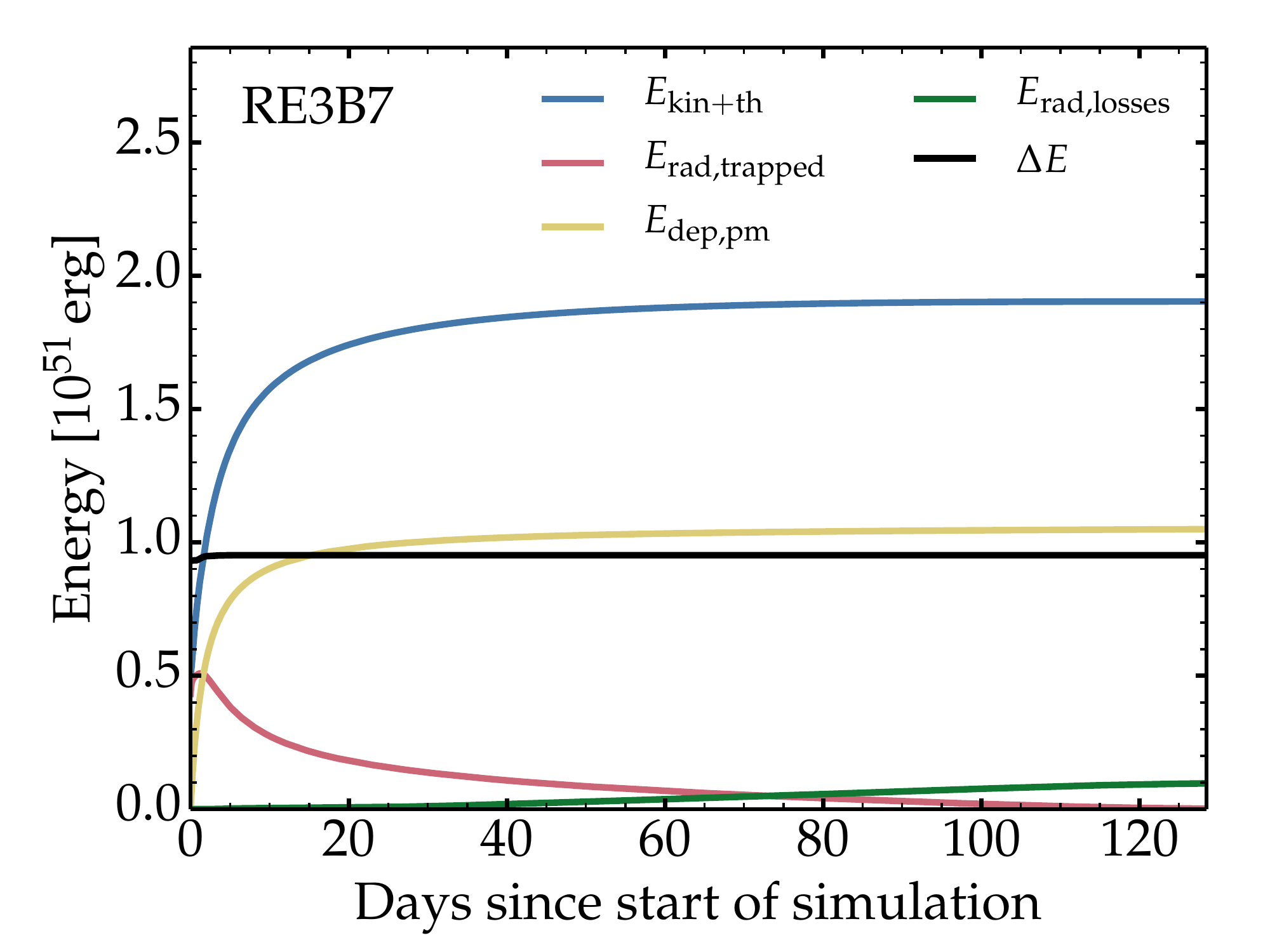}
\caption{Evolution of the various energy components in the
\heracles\ simulations. We show $E_{\rm kin+th}$, which
sums the kinetic and thermal (i.e., gas) energy;
$E_{\rm rad, trapped}$, which is the trapped radiation energy;
$E_{\rm dep, pm}$, which is the energy deposited by the magnetar;
$E_{\rm rad, losses}$, which is the radiative energy streaming out
through the outer grid boundary; and
$\Delta E$, which is defined as $E_{\rm kin+th}$ + $E_{\rm rad, trapped}$ +
$E_{\rm rad, losses}$ - $E_{\rm dep, pm}$ and should be constant.
The start of the simulation is 1.15\,d in all simulations --- any magnetar
energy deposited prior to that is not accounted for in the \heracles\
simulations.}
\label{fig_check_energy}
\end{figure*}

In this appendix we provide additional illustrations for the models
discussed in the main body of the paper.
In Fig.~\ref{fig_check_energy}, we discuss the evolution of
the various forms of energy on the grid, similarly to the top
panel of Fig.~\ref{fig_ener}.

\begin{figure*}
   \centering
       \includegraphics[width=0.49\hsize]{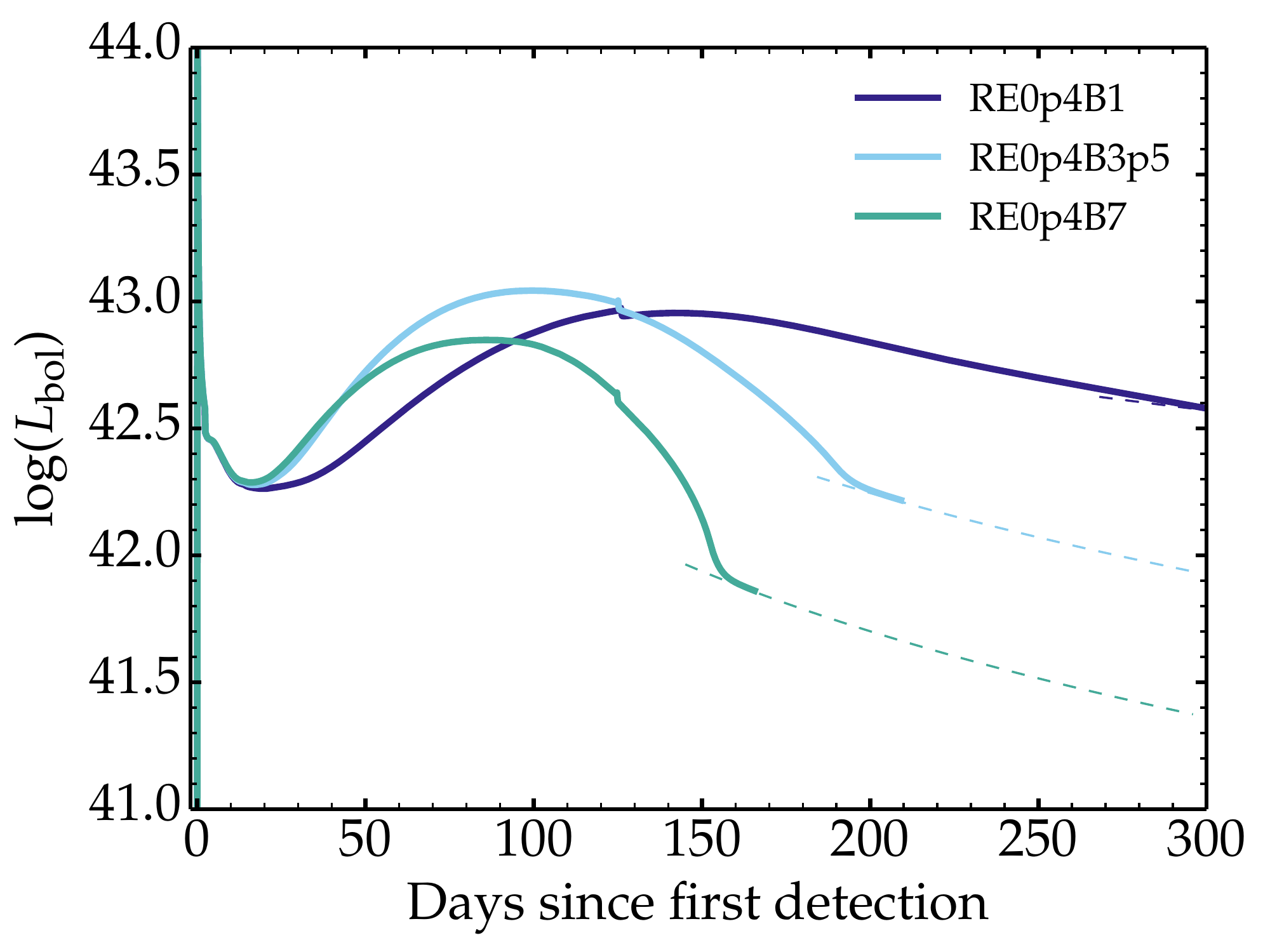}
       \includegraphics[width=0.49\hsize]{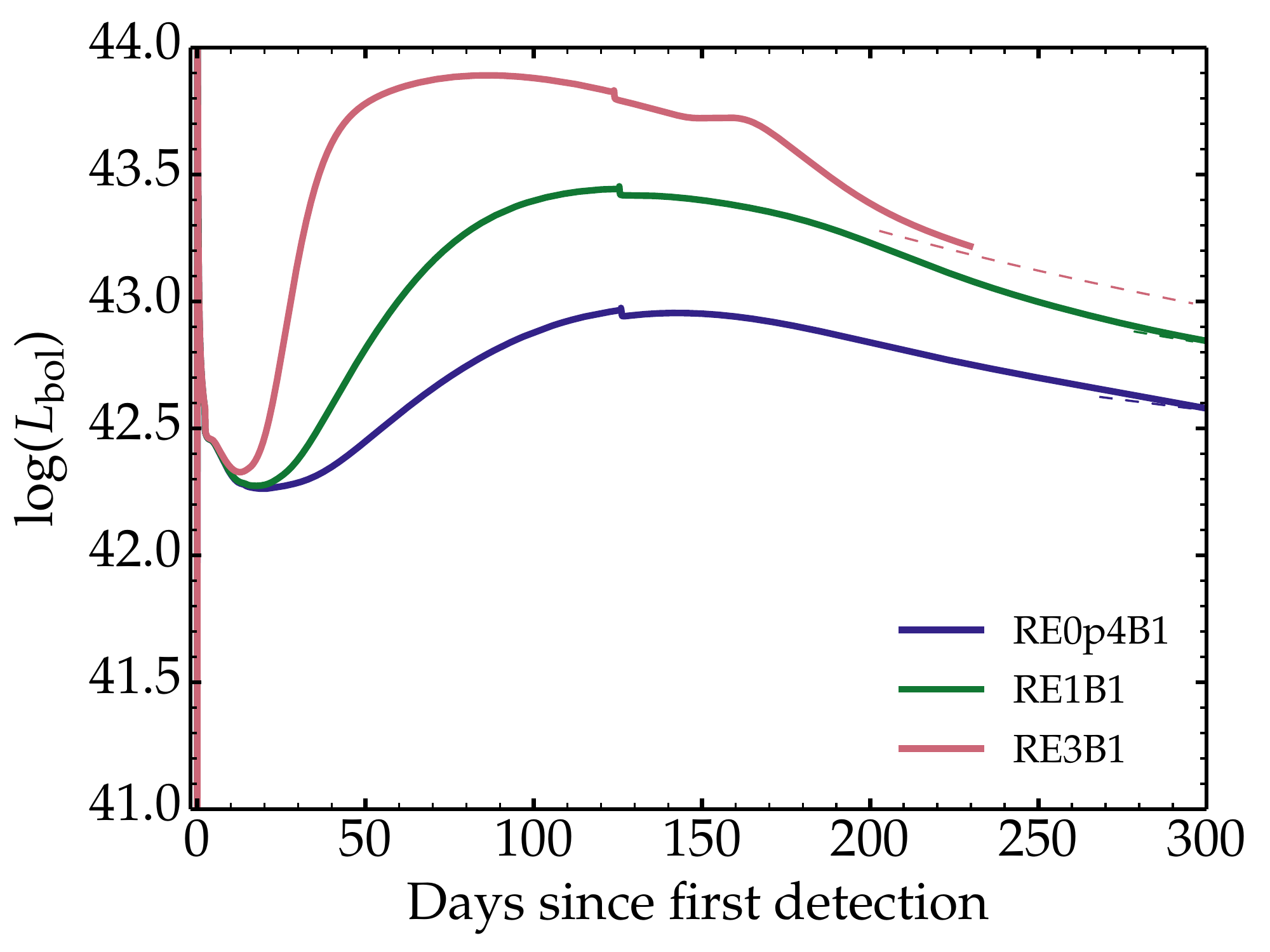}
       \includegraphics[width=0.49\hsize]{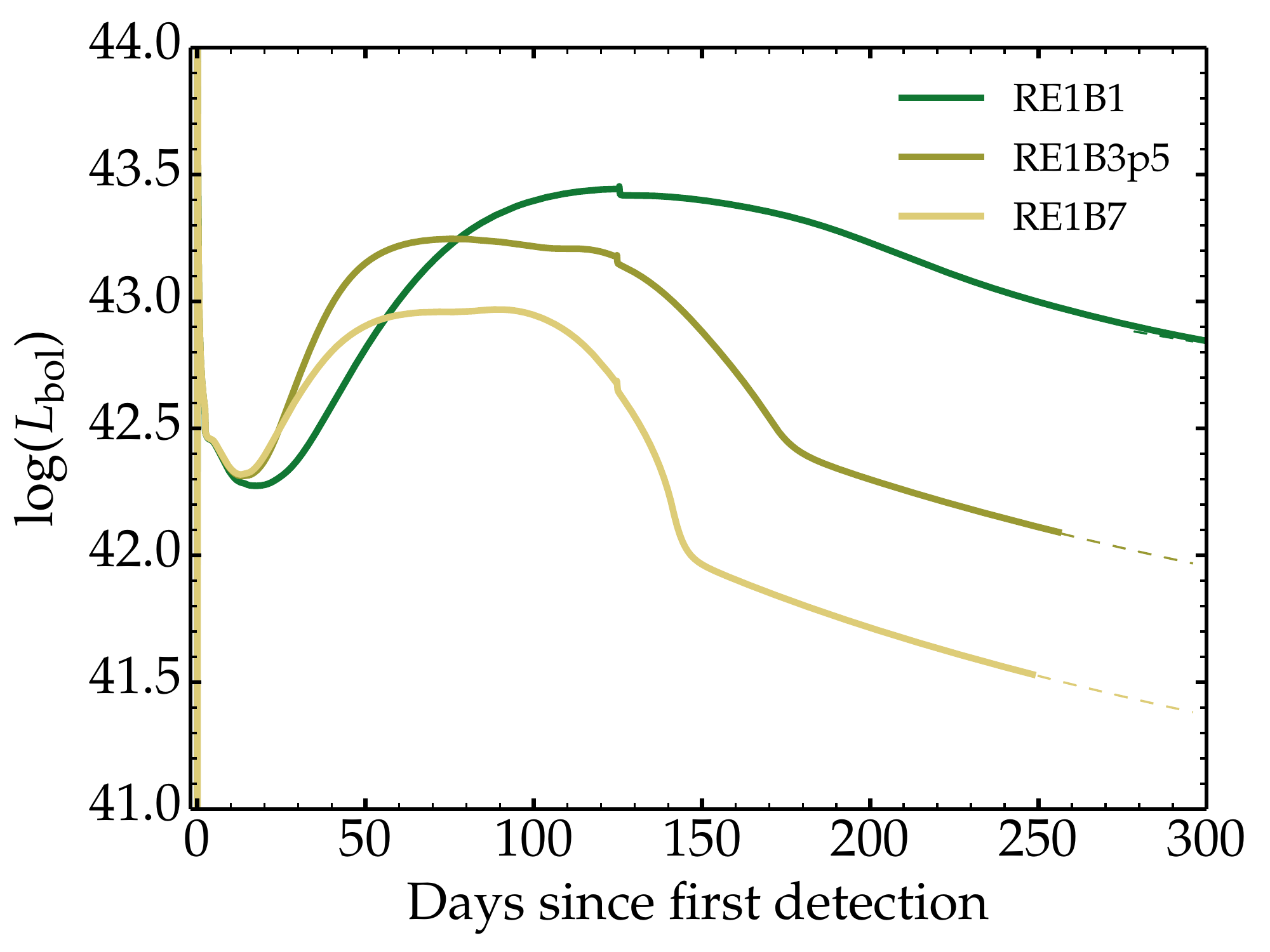}
       \includegraphics[width=0.49\hsize]{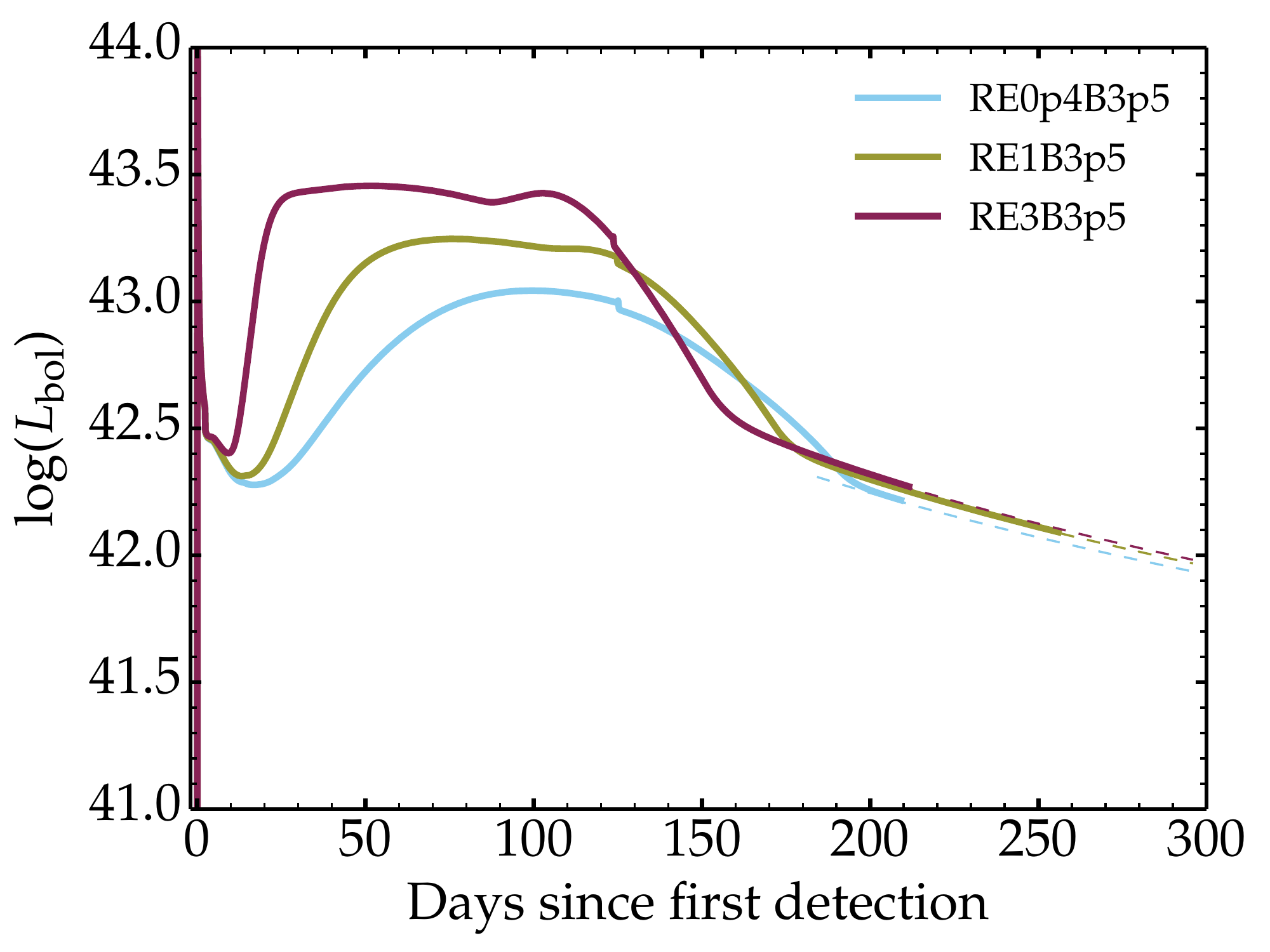}
       \includegraphics[width=0.49\hsize]{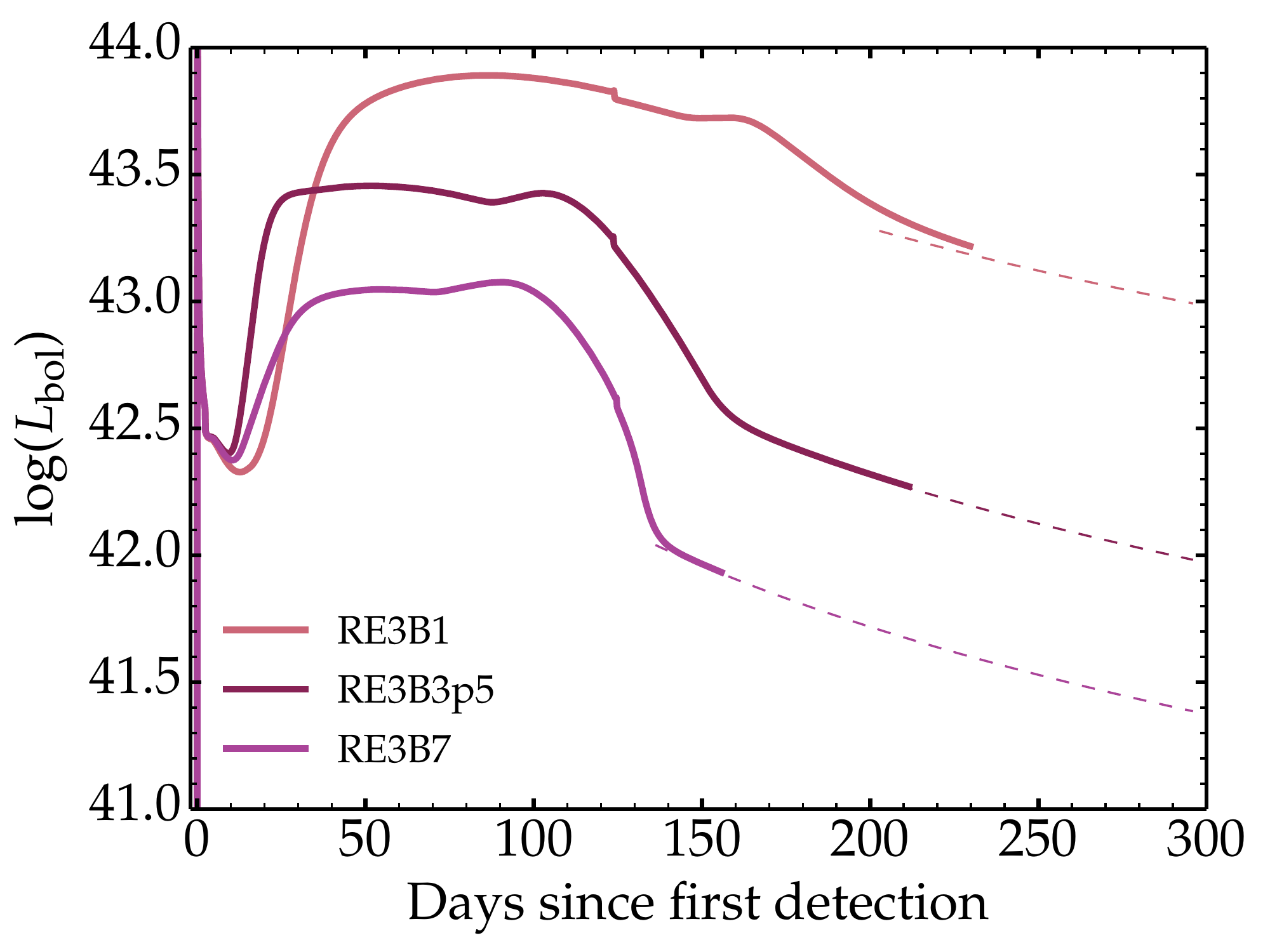}
       \includegraphics[width=0.49\hsize]{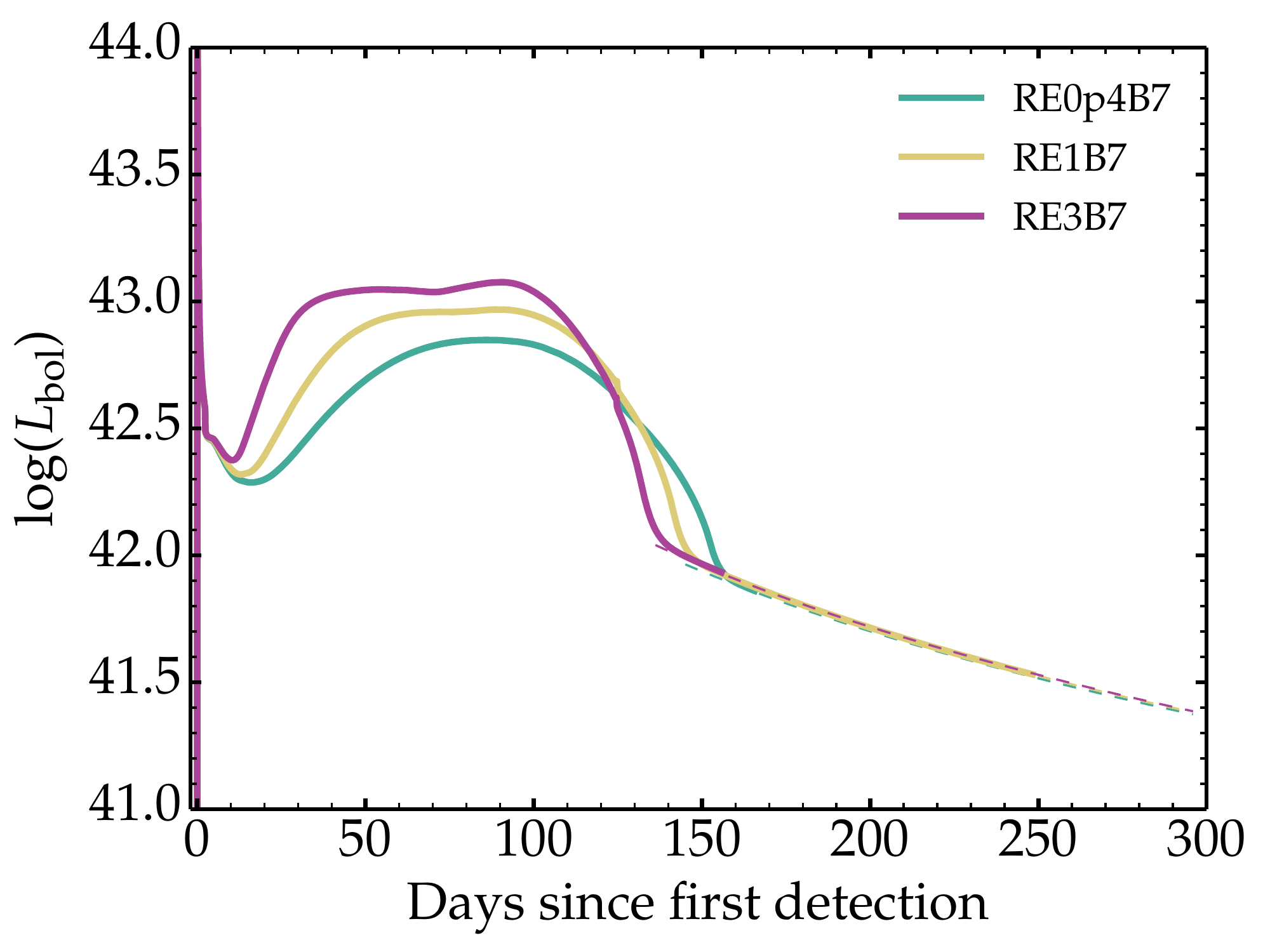}
   \caption{Bolometric light curves from \heracles\ simulations for the
   RSG progenitor. At $\sim$\,130\,d, the shock between ejecta and progenitor wind leaved the grid
   and causes a slight glitch in the luminosity. In some models (e.g., RE3B7), the light curve shows
   a small bump at late times in the photospheric phase.
   This feature is associated with the dense shell that results from the dynamical
   influence of the magnetar power (even though we adopt a broad energy deposition profile;
   see Section~\ref{sect_edep}).
   Models with a higher initial magnetar energy and/or a higher magnetar field tend
   to produce a more massive dense shell and show this late light curve bump.
   The glitch at 125\,d occurs when the outer shock with the progenitor wind leaves
   the Eulerian grid at 10$^{16}$\,cm.
}
   \label{fig_pm_prop}
    \end{figure*}

We also show in Fig.~\ref{fig_pm_prop} the bolometric light curves
for the magnetar-powered SN models discussed in Section~\ref{sect_grid}
but this time grouped by triads of the same magnetar initial rotational
energy (but different magnetic field) or the same magnetic field (but
different magnetar rotational energy).

\begin{thebibliography}{51}
\expandafter\ifx\csname natexlab\endcsname\relax\def\natexlab#1{#1}\fi

\bibitem[{{Arnett}(1982)}]{arnett_82}
{Arnett}, W.~D. 1982, \apj, 253, 785

\bibitem[{{Barkat} {et~al.}(1967){Barkat}, {Rakavy}, \&
  {Sack}}]{barkat_pisn_67}
{Barkat}, Z., {Rakavy}, G., \& {Sack}, N. 1967, Physical Review Letters, 18,
  379

\bibitem[{{Bersten} {et~al.}(2016){Bersten}, {Benvenuto}, {Orellana}, \&
  {Nomoto}}]{bersten_15lh_16}
{Bersten}, M.~C., {Benvenuto}, O.~G., {Orellana}, M., \& {Nomoto}, K. 2016,
  \apjl, 817, L8

\bibitem[{{Bucciantini} {et~al.}(2008){Bucciantini}, {Quataert}, {Arons},
  {Metzger}, \& {Thompson}}]{bucciantini+08}
{Bucciantini}, N., {Quataert}, E., {Arons}, J., {Metzger}, B.~D., \&
  {Thompson}, T.~A. 2008, \mnras, 383, L25

\bibitem[{{Bucciantini} {et~al.}(2009){Bucciantini}, {Quataert}, {Metzger},
  {Thompson}, {Arons}, \& {Del Zanna}}]{bucciantini_pm_09}
{Bucciantini}, N., {Quataert}, E., {Metzger}, B.~D., {et~al.} 2009, \mnras,
  396, 2038

\bibitem[{{Burrows} {et~al.}(2007){Burrows}, {Dessart}, {Livne}, {Ott}, \&
  {Murphy}}]{burrows_MHD_07}
{Burrows}, A., {Dessart}, L., {Livne}, E., {Ott}, C.~D., \& {Murphy}, J. 2007,
  \apj, 664, 416

\bibitem[{{Chen} {et~al.}(2016){Chen}, {Woosley}, \&
  {Sukhbold}}]{chen_pm_2d_16}
{Chen}, K.-J., {Woosley}, S.~E., \& {Sukhbold}, T. 2016, \apj, 832, 73

\bibitem[{{Chevalier} \& {Dwarkadas}(1995)}]{chevalier_dwarkadas_95}
{Chevalier}, R.~A. \& {Dwarkadas}, V.~V. 1995, \apjl, 452, L45

\bibitem[{{Dessart} {et~al.}(2015){Dessart}, {Audit}, \& {Hillier}}]{D15_2n}
{Dessart}, L., {Audit}, E., \& {Hillier}, D.~J. 2015, \mnras, 449, 4304

\bibitem[{{Dessart} \& {Hillier}(2005)}]{DH05a}
{Dessart}, L. \& {Hillier}, D.~J. 2005, \aap, 437, 667

\bibitem[{{Dessart} \& {Hillier}(2006)}]{DH06}
{Dessart}, L. \& {Hillier}, D.~J. 2006, \aap, 447, 691

\bibitem[{{Dessart} \& {Hillier}(2008)}]{D08_time}
{Dessart}, L. \& {Hillier}, D.~J. 2008, \mnras, 383, 57

\bibitem[{{Dessart} \& {Hillier}(2010)}]{DH10}
{Dessart}, L. \& {Hillier}, D.~J. 2010, \mnras, 405, 2141

\bibitem[{{Dessart} \& {Hillier}(2011)}]{DH11_2p}
{Dessart}, L. \& {Hillier}, D.~J. 2011, \mnras, 410, 1739

\bibitem[{{Dessart} {et~al.}(2016){Dessart}, {Hillier}, {Audit}, {Livne}, \&
  {Waldman}}]{D16_2n}
{Dessart}, L., {Hillier}, D.~J., {Audit}, E., {Livne}, E., \& {Waldman}, R.
  2016, \mnras, 458, 2094

\bibitem[{{Dessart} {et~al.}(2013{\natexlab{a}}){Dessart}, {Hillier},
  {Waldman}, \& {Livne}}]{d13_sn2p}
{Dessart}, L., {Hillier}, D.~J., {Waldman}, R., \& {Livne}, E.
  2013{\natexlab{a}}, \mnras, 433, 1745

\bibitem[{{Dessart} {et~al.}(2012){Dessart}, {Hillier}, {Waldman}, {Livne}, \&
  {Blondin}}]{d12_magnetar}
{Dessart}, L., {Hillier}, D.~J., {Waldman}, R., {Livne}, E., \& {Blondin}, S.
  2012, \mnras, 426, L76

\bibitem[{{Dessart} {et~al.}(2013{\natexlab{b}}){Dessart}, {Waldman}, {Livne},
  {Hillier}, \& {Blondin}}]{d13_pisn}
{Dessart}, L., {Waldman}, R., {Livne}, E., {Hillier}, D.~J., \& {Blondin}, S.
  2013{\natexlab{b}}, \mnras, 428, 3227

\bibitem[{{Duncan} \& {Thompson}(1992)}]{duncan_pm_92}
{Duncan}, R.~C. \& {Thompson}, C. 1992, \apjl, 392, L9

\bibitem[{{Folatelli} {et~al.}(2006){Folatelli}, {Contreras}, {Phillips},
  {Woosley}, {Blinnikov}, {Morrell}, {Suntzeff}, {Lee}, {Hamuy},
  {Gonz{\'a}lez}, {Krzeminski}, {Roth}, {Li}, {Filippenko}, {Foley},
  {Freedman}, {Madore}, {Persson}, {Murphy}, {Boissier}, {Galaz},
  {Gonz{\'a}lez}, {McCarthy}, {McWilliam}, \& {Pych}}]{folatelli_05bf_06}
{Folatelli}, G., {Contreras}, C., {Phillips}, M.~M., {et~al.} 2006, \apj, 641,
  1039

\bibitem[{{Gal-Yam} {et~al.}(2009){Gal-Yam}, {Mazzali}, {Ofek}, {Nugent},
  {Kulkarni}, {Kasliwal}, {Quimby}, {Filippenko}, {Cenko}, {Chornock},
  {Waldman}, {Kasen}, {Sullivan}, {Beshore}, {Drake}, {Thomas}, {Bloom},
  {Poznanski}, {Miller}, {Foley}, {Silverman}, {Arcavi}, {Ellis}, \&
  {Deng}}]{galyam_07bi_09}
{Gal-Yam}, A., {Mazzali}, P., {Ofek}, E.~O., {et~al.} 2009, \nat, 462, 624

\bibitem[{{Gezari} {et~al.}(2009){Gezari}, {Halpern}, {Grupe}, {Yuan},
  {Quimby}, {McKay}, {Chamarro}, {Sisson}, {Akerlof}, {Wheeler}, {Brown},
  {Cenko}, {Rau}, {Djordjevic}, \& {Terndrup}}]{gezari_08es_09}
{Gezari}, S., {Halpern}, J.~P., {Grupe}, D., {et~al.} 2009, \apj, 690, 1313

\bibitem[{{Gonz{\'a}lez} {et~al.}(2007){Gonz{\'a}lez}, {Audit}, \&
  {Huynh}}]{gonzalez_heracles_07}
{Gonz{\'a}lez}, M., {Audit}, E., \& {Huynh}, P. 2007, \aap, 464, 429

\bibitem[{{Guti{\'e}rrez} {et~al.}(2017){Guti{\'e}rrez}, {Anderson}, {Hamuy},
  {Morrell}, {Gonz{\'a}lez-Gaitan}, {Stritzinger}, {Phillips}, {Galbany},
  {Folatelli}, {Dessart}, {Contreras}, {Della Valle}, {Freedman}, {Hsiao},
  {Krisciunas}, {Madore}, {Maza}, {Suntzeff}, {Prieto}, {Gonz{\'a}lez},
  {Cappellaro}, {Navarrete}, {Pizzella}, {Ruiz}, {Smith}, \&
  {Turatto}}]{gutierrez_pap1_17}
{Guti{\'e}rrez}, C.~P., {Anderson}, J.~P., {Hamuy}, M., {et~al.} 2017, ArXiv
  e-prints [\eprint[arXiv]{1709.02487}]

\bibitem[{{Hamuy}(2003)}]{hamuy_03}
{Hamuy}, M. 2003, \apj, 582, 905

\bibitem[{{Hamuy} {et~al.}(1988){Hamuy}, {Suntzeff}, {Gonzalez}, \&
  {Martin}}]{hamuy_87A_88}
{Hamuy}, M., {Suntzeff}, N.~B., {Gonzalez}, R., \& {Martin}, G. 1988, \aj, 95,
  63

\bibitem[{{Inserra} {et~al.}(2013){Inserra}, {Smartt}, {Jerkstrand}, {Valenti},
  {Fraser}, {Wright}, {Smith}, {Chen}, {Kotak}, {Pastorello}, {Nicholl},
  {Bresolin}, {Kudritzki}, {Benetti}, {Botticella}, {Burgett}, {Chambers},
  {Ergon}, {Flewelling}, {Fynbo}, {Geier}, {Hodapp}, {Howell}, {Huber},
  {Kaiser}, {Leloudas}, {Magill}, {Magnier}, {McCrum}, {Metcalfe}, {Price},
  {Rest}, {Sollerman}, {Sweeney}, {Taddia}, {Taubenberger}, {Tonry},
  {Wainscoat}, {Waters}, \& {Young}}]{inserra_slsn_13}
{Inserra}, C., {Smartt}, S.~J., {Jerkstrand}, A., {et~al.} 2013, \apj, 770, 128

\bibitem[{{Jerkstrand} {et~al.}(2017){Jerkstrand}, {Smartt}, {Inserra},
  {Nicholl}, {Chen}, {Kr{\"u}hler}, {Sollerman}, {Taubenberger}, {Gal-Yam},
  {Kankare}, {Maguire}, {Fraser}, {Valenti}, {Sullivan}, {Cartier}, \&
  {Young}}]{jerkstrand_slsnic_17}
{Jerkstrand}, A., {Smartt}, S.~J., {Inserra}, C., {et~al.} 2017, \apj, 835, 13

\bibitem[{{Kasen} \& {Bildsten}(2010)}]{KB10}
{Kasen}, D. \& {Bildsten}, L. 2010, \apj, 717, 245

\bibitem[{{Maeda} {et~al.}(2007){Maeda}, {Tanaka}, {Nomoto}, {Tominaga},
  {Kawabata}, {Mazzali}, {Umeda}, {Suzuki}, \& {Hattori}}]{maeda_05bf_07}
{Maeda}, K., {Tanaka}, M., {Nomoto}, K., {et~al.} 2007, \apj, 666, 1069

\bibitem[{{Metzger} {et~al.}(2011){Metzger}, {Giannios}, {Thompson},
  {Bucciantini}, \& {Quataert}}]{metzger_pm_11}
{Metzger}, B.~D., {Giannios}, D., {Thompson}, T.~A., {Bucciantini}, N., \&
  {Quataert}, E. 2011, \mnras, 413, 2031

\bibitem[{{Metzger} {et~al.}(2015){Metzger}, {Margalit}, {Kasen}, \&
  {Quataert}}]{metzger_pm_15}
{Metzger}, B.~D., {Margalit}, B., {Kasen}, D., \& {Quataert}, E. 2015, \mnras,
  454, 3311

\bibitem[{{Miller} {et~al.}(2009){Miller}, {Chornock}, {Perley},
  {Ganeshalingam}, {Li}, {Butler}, {Bloom}, {Smith}, {Modjaz}, {Poznanski},
  {Filippenko}, {Griffith}, {Shiode}, \& {Silverman}}]{miller_08es_09}
{Miller}, A.~A., {Chornock}, R., {Perley}, D.~A., {et~al.} 2009, \apj, 690,
  1303

\bibitem[{{M{\"o}sta} {et~al.}(2015){M{\"o}sta}, {Ott}, {Radice}, {Roberts},
  {Schnetter}, \& {Haas}}]{moesta_15}
{M{\"o}sta}, P., {Ott}, C.~D., {Radice}, D., {et~al.} 2015, \nat, 528, 376

\bibitem[{{Nicholl} {et~al.}(2014){Nicholl}, {Smartt}, {Jerkstrand}, {Inserra},
  {Anderson}, {Baltay}, {Benetti}, {Chen}, {Elias-Rosa}, {Feindt}, {Fraser},
  {Gal-Yam}, {Hadjiyska}, {Howell}, {Kotak}, {Lawrence}, {Leloudas},
  {Margheim}, {Mattila}, {McCrum}, {McKinnon}, {Mead}, {Nugent}, {Rabinowitz},
  {Rest}, {Smith}, {Sollerman}, {Sullivan}, {Taddia}, {Valenti}, {Walker}, \&
  {Young}}]{nicholl_slsn_14}
{Nicholl}, M., {Smartt}, S.~J., {Jerkstrand}, A., {et~al.} 2014, \mnras, 444,
  2096

\bibitem[{{Nicholl} {et~al.}(2013){Nicholl}, {Smartt}, {Jerkstrand}, {Inserra},
  {McCrum}, {Kotak}, {Fraser}, {Wright}, {Chen}, {Smith}, {Young}, {Sim},
  {Valenti}, {Howell}, {Bresolin}, {Kudritzki}, {Tonry}, {Huber}, {Rest},
  {Pastorello}, {Tomasella}, {Cappellaro}, {Benetti}, {Mattila}, {Kankare},
  {Kangas}, {Leloudas}, {Sollerman}, {Taddia}, {Berger}, {Chornock}, {Narayan},
  {Stubbs}, {Foley}, {Lunnan}, {Soderberg}, {Sanders}, {Milisavljevic},
  {Margutti}, {Kirshner}, {Elias-Rosa}, {Morales-Garoffolo}, {Taubenberger},
  {Botticella}, {Gezari}, {Urata}, {Rodney}, {Riess}, {Scolnic}, {Wood-Vasey},
  {Burgett}, {Chambers}, {Flewelling}, {Magnier}, {Kaiser}, {Metcalfe},
  {Morgan}, {Price}, {Sweeney}, \& {Waters}}]{nicholl_slsn_13}
{Nicholl}, M., {Smartt}, S.~J., {Jerkstrand}, A., {et~al.} 2013, \nat, 502, 346

\bibitem[{{Ofek} {et~al.}(2007){Ofek}, {Cameron}, {Kasliwal}, {Gal-Yam}, {Rau},
  {Kulkarni}, {Frail}, {Chandra}, {Cenko}, {Soderberg}, \&
  {Immler}}]{ofek_06gy_07}
{Ofek}, E.~O., {Cameron}, P.~B., {Kasliwal}, M.~M., {et~al.} 2007, \apjl, 659,
  L13

\bibitem[{{Polshaw} {et~al.}(2016){Polshaw}, {Kotak}, {Dessart}, {Fraser},
  {Gal-Yam}, {Inserra}, {Sim}, {Smartt}, {Sollerman}, {Baltay}, {Rabinowitz},
  {Benetti}, {Botticella}, {Campbell}, {Chen}, {Galbany}, {McKinnon},
  {Nicholl}, {Smith}, {Sullivan}, {Tak{\'a}ts}, {Valenti}, \&
  {Young}}]{polshaw_lsq13fn_16}
{Polshaw}, J., {Kotak}, R., {Dessart}, L., {et~al.} 2016, \aap, 588, A1

\bibitem[{{Smith} {et~al.}(2007){Smith}, {Li}, {Foley}, {Wheeler}, {Pooley},
  {Chornock}, {Filippenko}, {Silverman}, {Quimby}, {Bloom}, \&
  {Hansen}}]{smith_06gy_07}
{Smith}, N., {Li}, W., {Foley}, R.~J., {et~al.} 2007, \apj, 666, 1116

\bibitem[{{Stoll} {et~al.}(2011){Stoll}, {Prieto}, {Stanek}, {Pogge},
  {Szczygie{\l}}, {Pojma{\'n}ski}, {Antognini}, \& {Yan}}]{stoll_10jl_11}
{Stoll}, R., {Prieto}, J.~L., {Stanek}, K.~Z., {et~al.} 2011, \apj, 730, 34

\bibitem[{{Sukhbold} \& {Woosley}(2016)}]{sukhbold_woosley_16}
{Sukhbold}, T. \& {Woosley}, S.~E. 2016, \apjl, 820, L38

\bibitem[{{Suzuki} \& {Maeda}(2017)}]{suzuki_pm_2d_17}
{Suzuki}, A. \& {Maeda}, K. 2017, \mnras, 466, 2633

\bibitem[{{Taddia} {et~al.}(2017){Taddia}, {Sollerman}, {Fremling},
  {Karamehmetoglu}, {Quimby}, {Gal-Yam}, {Yaron}, {Kasliwal}, {Kulkarni},
  {Nugent}, {Smadja}, \& {Tao}}]{taddia_pm_17}
{Taddia}, F., {Sollerman}, J., {Fremling}, C., {et~al.} 2017, ArXiv:1709.08386
  [\eprint[arXiv]{1709.08386}]

\bibitem[{{Taddia} {et~al.}(2012){Taddia}, {Stritzinger}, {Sollerman},
  {Phillips}, {Anderson}, {Ergon}, {Folatelli}, {Fransson}, {Freedman},
  {Hamuy}, {Morrell}, {Pastorello}, {Persson}, \& {Gonzalez}}]{taddia_2pec_12}
{Taddia}, F., {Stritzinger}, M.~D., {Sollerman}, J., {et~al.} 2012, \aap, 537,
  A140

\bibitem[{{Terreran} {et~al.}(2017){Terreran}, {Pumo}, {Chen}, {Moriya},
  {Taddia}, {Dessart}, {Zampieri}, {Smartt}, {Benetti}, {Inserra},
  {Cappellaro}, {Nicholl}, {Fraser}, {Wyrzykowski}, {Udalski}, {Howell},
  {McCully}, {Valenti}, {Dimitriadis}, {Maguire}, {Sullivan}, {Smith}, {Yaron},
  {Young}, {Anderson}, {Della Valle}, {Elias-Rosa}, {Gal-Yam}, {Jerkstrand},
  {Kankare}, {Pastorello}, {Sollerman}, {Turatto}, {Kostrzewa-Rutkowska},
  {Koz{\l}owski}, {Mr{\'o}z}, {Pawlak}, {Pietrukowicz}, {Poleski}, {Skowron},
  {Skowron}, {Soszy{\'n}ski}, {Szyma{\'n}ski}, \&
  {Ulaczyk}}]{terreran_slsn2_17}
{Terreran}, G., {Pumo}, M.~L., {Chen}, T.-W., {et~al.} 2017, Nature Astronomy,
  1, 228

\bibitem[{{Thompson} {et~al.}(2004){Thompson}, {Chang}, \&
  {Quataert}}]{thompson_pm_04}
{Thompson}, T.~A., {Chang}, P., \& {Quataert}, E. 2004, \apj, 611, 380

\bibitem[{{Usov}(1992)}]{usov_pm_92}
{Usov}, V.~V. 1992, \nat, 357, 472

\bibitem[{{Utrobin} \& {Chugai}(2005)}]{UC05}
{Utrobin}, V.~P. \& {Chugai}, N.~N. 2005, \aap, 441, 271

\bibitem[{{Utrobin} {et~al.}(2017){Utrobin}, {Wongwathanarat}, {Janka}, \&
  {M{\"u}ller}}]{utrobin_2p_17}
{Utrobin}, V.~P., {Wongwathanarat}, A., {Janka}, H.-T., \& {M{\"u}ller}, E.
  2017, \apj, 846, 37

\bibitem[{{Vaytet} {et~al.}(2011){Vaytet}, {Audit}, {Dubroca}, \&
  {Delahaye}}]{vaytet_mg_11}
{Vaytet}, N.~M.~H., {Audit}, E., {Dubroca}, B., \& {Delahaye}, F. 2011, \jqsrt,
  112, 1323

\bibitem[{{Woosley}(2010)}]{woosley_pm_10}
{Woosley}, S.~E. 2010, \apjl, 719, L204

\end{thebibliography}
\end{document}